\documentclass[traditabstract]{aa_katsuta}
\usepackage{txfonts}
\usepackage{graphicx}
\usepackage{color}

\usepackage{natbib,twoopt}
 \usepackage[breaklinks=true]{hyperref} 
 \bibpunct{(}{)}{;}{a}{}{,}    
 \newcommandtwoopt{\citeads}[3][][]{\href{http://adsabs.harvard.edu/abs/#3}%
                                        {\citealp[#1][#2]{#3}}}
 \newcommandtwoopt{\citepads}[3][][]{\href{http://adsabs.harvard.edu/abs/#3}%
                                        {\citep[#1][#2]{#3}}}
 \newcommandtwoopt{\citetads}[3][][]{\href{http://adsabs.harvard.edu/abs/#3}%
                                        {\citet[#1][#2]{#3}}}
 \newcommandtwoopt{\citeyearads}[3][][]%
   {\href{http://adsabs.harvard.edu/abs/#3}{\citeyear[#1][#2]{#3}}}
   
\def\FL{\emph{Fermi}-LAT}
\def\SZ{\emph{Suzaku}}
\def\AS{\emph{ASCA}}
\def\CB{Centaurus\,B}
\def\MOD{\rm}
\def\JKC{black}

\begin{document}

\title{\FL\ and \SZ\ Observations of the Radio Galaxy \CB}

\author{
J.~Katsuta \inst{2,1}
\and
Y.~T.~Tanaka \inst{3,1}
\and
{\L}.~Stawarz \inst{4,5}
\and
S.~P.~O'Sullivan \inst{6}
\and
C.~C.~Cheung \inst{7}
\and
J.~Kataoka \inst{8}
\and
S.~Funk \inst{2}
\and
T.~Yuasa \inst{4}
\and
H.~Odaka \inst{4}
\and
T.~Takahashi \inst{4}
\and
J.~Svoboda \inst{9}
}

\institute{Corresponding author: J.~Katsuta, katsuta@slac.stanford.edu; Y.~Tanaka, ytanaka@hep01.hepl.hiroshima-u.ac.jp
\and
W. W. Hansen Experimental Physics Laboratory, Kavli Institute for Particle Astrophysics and Cosmology, Department of Physics and SLAC National Accelerator Laboratory, Stanford University, Stanford, CA 94305, USA
\and
Hiroshima Astrophysical Science Center, Hiroshima University, 1-3-1 Kagamiyama, Higashi-Hiroshima, Hiroshima 739-8526, Japan
\and
Institute of Space and Astronautical Science, Japan Aerospace Exploration Agency, 3-1-1 Yoshinodai, Chuo-ku, Sagamihara, Kanagawa 252-5210, Japan
\and
Astronomical Observatory, Jagiellonian University, ul. Orla 171, 30-244 Krak\'ow, Poland
\and
Sydney Institute for Astronomy, School of Physics A28, University of Sydney, NSW 2006, Australia
\and
National Research Council Research Associate, National Academy of Sciences, Washington, DC 20001, resident at Naval Research Laboratory, Washington, DC 20375, USA
\and
Research Institute for Science and Engineering, Waseda University, 3-4-1, Okubo, Shinjuku, Tokyo 169-8555, Japan
\and
European Space Astronomy Centre of ESA, PO Box 78, Villanueva de la Ca\~{n}ada, 28691 Madrid, Spain
}

\abstract{
\CB\ is a nearby radio galaxy positioned in the Southern hemisphere close to the Galactic plane. 
\textcolor{\JKC}{\MOD Here we present a detailed analysis of about 43 months of accumulated \FL\ data of the $\gamma$-ray counterpart of the source initially reported in the 2nd \FL\ catalog, and of newly acquired \SZ\ X-ray data. We confirm its detection at GeV photon energies, and analyze the extension and variability of the $\gamma$-ray source in the LAT dataset, in which it appears as a steady $\gamma$-ray emitter.}
The X-ray core of \CB\ is detected as a bright source of a continuum radiation. 
We do not detect however any diffuse X-ray emission from the known radio lobes, with the provided upper limit only marginally consistent with the previously claimed \emph{ASCA} flux. 
Two scenarios that connect the X-ray and $\gamma$-ray properties are considered.
In the first one, we assume that the diffuse non-thermal X-ray emission component is not significantly below the derived \SZ\ upper limit. In this case, modeling the inverse-Compton emission shows that the observed $\gamma$-ray flux of the source may in principle be produced within the lobes. This association would imply that efficient \emph{in-situ} acceleration of the radiating electrons is occurring and that the lobes are dominated by the pressure from the relativistic particles. 
In the second scenario, with the diffuse X-ray emission well below the \SZ\ upper limits, the lobes in the system are instead dominated by the magnetic pressure.
In this case, the observed $\gamma$-ray flux is not likely to be produced within the lobes, but instead within the nuclear parts of the jet. By means of synchrotron self-Compton modeling we show that this possibility could be consistent with the broad-band data collected for the unresolved core of \CB, including the newly derived \SZ\ spectrum.}

\maketitle

\section{Introduction}
\label{sec: intro}

\CB\ (PKS\,1343$-$601) is a relatively nearby \citep[$z=0.0129$,][distance $D = 56$\,Mpc]{west89} radio-loud active galaxy. 
Despite its high radio flux --- \CB\ is in fact the fifth brightest radio galaxy in the sky \citep{JLM01} --- it is relatively poorly studied. 
This is mainly due to both its proximity to the Galactic plane and the fact that it is not accessible from the Northern hemisphere. 
Initial studies suggested \CB\ was surrounded by a rich cluster of galaxies hidden behind the Milky Way and related to one of the largest concentrations of mass in the local Universe, the Great Attractor \citep{GA,kraan99,kraan00}. 
In addition, \CB\ was found close to the arrival detection of the ultra-high energy cosmic ray (UHECR) event detected recently by the P. Auger Observatory \citep{PAO,nagar08,moskalenko09}. 
It is currently not clear whether a positional coincidence between nearby active galactic nuclei (AGN) and the arrival directions of UHECRs robustly identifies those AGN as UHECR sources, or implies instead that nearby AGN simply trace the mass distribution in the Universe \citep[see][]{moskalenko09}.

Searches for the members of the galaxy cluster surrounding \CB\ at optical wavelengths are significantly hampered by a heavy foreground extinction in the direction of the source. Near-infrared surveys showed an enhancement of galaxies around the discussed object, but not as significant as expected, and in addition consisting of mostly late-type systems \citep{nagayama04,kraan05,schroeder07}. Hence it seems unlikely that \CB\ resides in a particularly massive cluster, although a somewhat less rich environment like a poor cluster or a galaxy group remain valid options. This conclusion is in agreement with the X-ray studies presented in the literature, which did not reveal a presence of any extended atmosphere of hot ionized gas surrounding the source. In particular, the \emph{Einstein} Observatory detected only an X-ray point source coincident with the radio core of \CB. 
Deep observations with the \AS\ satellite confirmed the \emph{Einstein} detection, and in addition provided some evidence for a low surface-brightness diffuse X-ray halo around the core \citep{tashiro98}. The total luminosity of this halo, $L_{\rm 2-10\,keV} \simeq 3.7 \times 10^{41}$\,erg\,s$^{-1}$, which was proposed to be associated with the extended radio lobes in the system, is however orders of magnitude below that expected for a rich cluster \citep[see the discussion in][]{JLM01}. 
Using the \emph{ROSAT} All Sky Survey data, \citet{ebeling02} did not find any X-ray feature, neither point-like nor extended, at the source position, and thus excluded large amounts of X-ray emitting gas analogous to those found in rich clusters.

A comprehensive study of \CB\ at radio wavelengths was presented by \citet{JLM01}. The compiled maps revealed prominent jets surrounded by diffuse lobes. 
The largest angular size of the radio structure is $\simeq 24'$, corresponding to a projected size of about $\simeq 380$\,kpc for a conversion scale of $265$\,pc\ arcsec$^{-1}$. 
The actual physical extension of the lobes in \CB\ may be larger than this, because of a likely intermediate inclination to the line of sight implied by the jet-to-counterjet brightness and polarization asymmetry. Both the large-scale radio morphology of the source, as well as its total radio power ($L_{\rm 1.4\,GHz} \simeq 5 \times 10^{41}$\,erg\,s$^{-1}$), suggests a classification of the object as an intermediate case between Fanaroff-Riley class\,I (FR\,I) and class\,II (FR\,II).

As noted before, \citet{tashiro98} claimed the X-ray detection of the core and the tentative discovery of an extended halo coinciding with the radio lobes of \CB. They argued that the most likely origin of the diffuse X-ray halo, best fitted by a power-law model with the photon index $\Gamma_{\rm X} \simeq 1.88 \pm 0.19$, is inverse-Compton emission of the non-thermal electrons within the lobes interacting with the cosmic microwave background (CMB) photon field. 
{\MOD This detection, in fact only the second of its kind following the case of Fornax\,A \citep{feigelson95,kaneda95}, was followed by a number of analogous studies targeting other radio galaxies with the XMM-\emph{Newton}, \SZ\ and \emph{Chandra} X-ray satellites \citep[and references therein]{kataoka05,croston05,isobe11}. }
For \CB, the \emph{Chandra} observations resolved the inner parts of the main (SW) jet at arcsec scale thanks to its excellent spatial resolution \citep{marshall05}. 
{\MOD The detection of the SW jet on slightly larger scales was also reported in the preliminary analysis of the XMM data presented in \citet{tashiro05}.}

Here we present a detailed analysis of \CB\ at $\gamma$-ray frequencies with the Large Area Telescope (LAT) onboard the \emph{Fermi} satellite, together with an X-ray analysis of the \SZ\ data we have acquired very recently. 
\CB\ is associated with the 2FGL J1346.5$-$6027 object in the \FL\ Second Source Catalog \citep[2FGL;][]{2LAC,2FGL}.
In our investigation we focus on the extended lobes in this system, since their high integrated radio flux (dominating over that of the unresolved core by about an order of magnitude) and exceptional extension on the sky (about half of a degree) suggest that \CB\ could be one of the few nearby radio galaxies for which the diffuse lobes may be detected and resolved, or at least viewed as extended, in $\gamma$-rays. 
The motivation of the study came from the robust identification of several other radio galaxies with \FL\ sources \citep{PerA,M87,MAGN,BLRG}, including the case of Centaurus\,A for which the 600\,kpc-scale lobes have in fact been successfully resolved at GeV photon energies {\citep{CenALobes,CenACore}. 
{\MOD These studies have provided important} clues for understanding energy dissipation processes within the extended lobes of radio-loud AGN in general, and particle acceleration mechanisms related to the production of UHECRs in particular \citep[see in this context][]{hardcastle09,osullivan09,peer12}.

\section{\FL\ Data}
\label{sec: obs}

The \emph{Fermi} Gamma-ray Space Telescope was launched in 2008 June 11. The LAT onboard the \emph{Fermi} satellite is a pair conversion telescope, equipped with solid state silicon trackers and cesium iodide calorimeters, sensitive to photons in a very broad energy band from $\sim 20$\,MeV to $\gtrsim 300$\,GeV. The LAT has a large effective area ($\sim 8000$\,cm$^2$ above 1\,GeV for on-axis events), instantaneously viewing $\simeq 2.4$\,sr of the sky with a good angular resolution (68\% containment radius better than $\sim 1^{\circ}$ above 1\,GeV). The tracker of the LAT is divided into \emph{front} and \emph{back} sections. The front section composed of thin converters has a better angular resolution (68\% containment radius $\simeq 0.7^{\circ}$ at 1\,GeV) than that of the back section (68\% containment radius $\simeq 1.2^{\circ}$ at 1\,GeV) composed of thicker converters. Details of the LAT instrument and data reduction are described in \citet{LAT}.

The LAT data used here were collected during $\sim 43$ months from 2008 August 4 to 2012 February 18. The \emph{source} event class was chosen and photons beyond the Earth zenith angle of $100^\circ$ were excluded to reduce the background from the Earth limb. 
We also applied the ROI-based zenith angle cut which excludes times where the region of interest intersects a given zenith angle cut.
The \textsc{P7SOURCE\_V6} instrument response functions were used for the analysis presented in this paper.

We utilized \textsc{gtlike} in the Science Tools analysis package\footnote{available at the \emph{Fermi} Science Support Center, http://fermi.gsfc.nasa.gov/ssc} for spectral fits. With \textsc{gtlike}, a binned maximum likelihood fit is performed on the spatial and spectral distributions of the observed $\gamma$-rays to optimize spectral parameters of the input model taking into account the energy dependence of the point-spread function (PSF). Analysis using \textsc{gtlike} is done in a $20^\circ\times20^\circ$ region around \CB, which is referred to as a region of interest (ROI). The fitting model includes other point sources whose positions are fixed at the values provided in the 2FGL catalog. The Galactic diffuse emission is modeled by \textsc{gal\_2yearp7v6\_v0.fits} while an isotropic component (instrumental and extragalactic diffuse backgrounds) by \textsc{isotropic\_p7v6source.txt}. Both background models are the standard diffuse emission models released by the LAT team. In each \textsc{gtlike} run, all point sources within the ROI and diffuse components in the model are fitted with their normalizations being free. In the analysis we also include 2FGL sources outside the ROI but within $15^\circ$ of \CB, with their parameters fixed at those given in the 2FGL catalog.

\subsection{Detection and Source Localization}
\label{sec: det}

\begin{figure*}
\begin{center}
\includegraphics[width=7.0in]{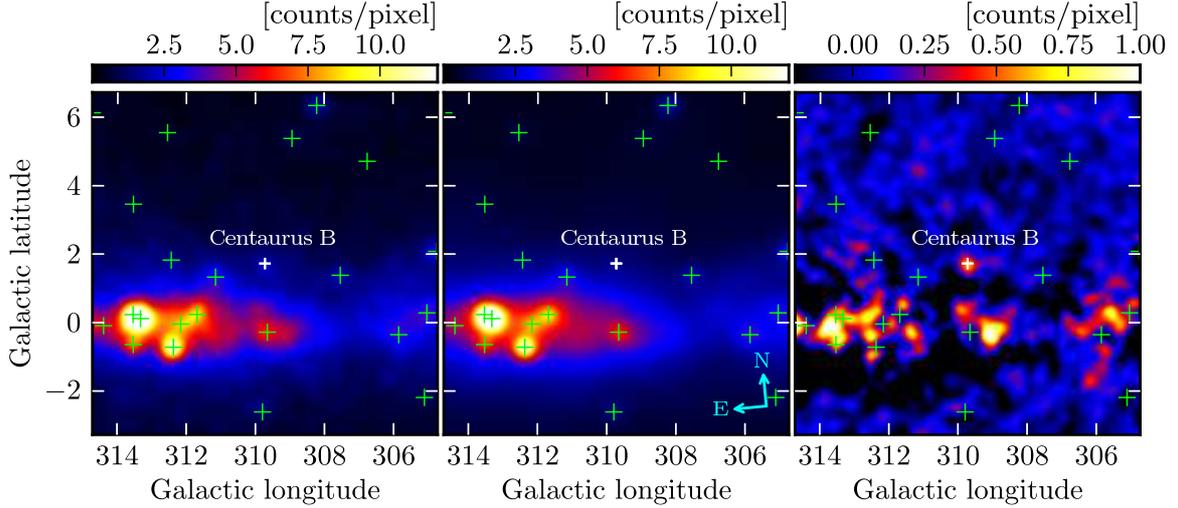}
\caption{\small
(Left) \FL\ count map using \emph{front} events above 1\,GeV around \CB\ in units of counts per pixel. The pixel size is $0\fdg 05$. Smoothing with a  Gaussian kernel of $\sigma =0 \fdg 15$ is applied. The background 2FGL sources contained in the region are shown as green crosses. The position of \CB\ is indicated as a white cross. (Middle) Background model map. (Right) Background-subtracted count map. 
\label{fig: cmap_wide}}
\end{center}
\end{figure*}

Figure~\ref{fig: cmap_wide} shows a smoothed count map of a $10^\circ\times10^\circ$ region around \CB\ above 1\,GeV, a corresponding background model map, and the background-subtracted count map. Here only \emph{front} events are used to achieve \textcolor{\JKC}{\MOD the best} angular resolution. The map includes contributions from the Galactic diffuse emission, isotropic diffuse background, and nearby discrete sources. The model parameters of diffuse components and nearby sources are set at the best-fit values obtained by \textsc{gtlike} using the data above 1\,GeV. In this fitting model, the 2FGL source associated with \CB\ (2FGL\,J1346.6$-$6027) is not included. The $\gamma$-ray source associated with \CB\ is visible in the background-subtracted map. The figure also shows residuals on the Galactic plane. These may be attributed to the imperfection of the Galactic diffuse model and/or the contributions from discrete sources not resolved from background, since the residuals are seen exclusively on the Galactic plane where the Galactic diffuse emission components have high intensities and complex spatial structures. 
This interpretation is also supported by the fact that all bright residuals are diffuse and/or accompanied by nearby 2FGL sources.
On the other hand, Figure~\ref{fig: cmap_wide} indicates that the source associated with \CB\ is a discrete feature in the residual map. Actually, the region of \CB\ is relatively off the plane\textcolor{\JKC}{\MOD : $(l,b) = (309\fdg 7, 1\fdg 7),$} where the Galactic diffuse emission is almost uniform with Galactic longitude and decreases monotonically with Galactic latitude. 

To study this $\gamma$-ray source in detail,
we generated a Test Statistic (TS) map around \CB. TS is defined as $-2\Delta\ln ({\rm likelihood})$ obtained by \textsc{gtlike} between models of the null hypothesis and an alternative. In this paper, we refer to TS as a comparison between models without a target source (null hypothesis) and with the source (alternative hypothesis) unless otherwise mentioned. In the presented TS map, TS at each grid position is calculated by placing a point source with a power-law energy distribution with photon index set free. 

\begin{figure}[!h] 
\begin{center}
\includegraphics[width=3.5in]{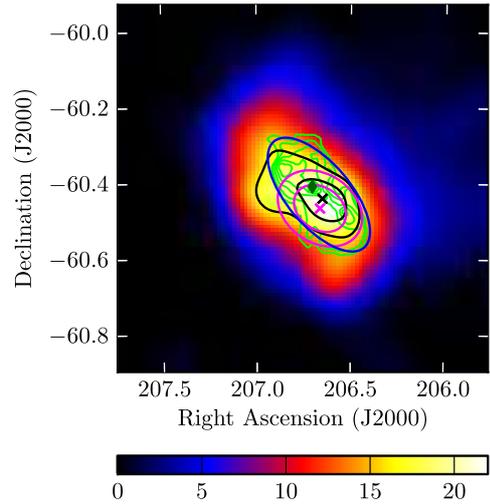}
\caption{\small \FL\ TS map obtained with a maximum likelihood analysis using \emph{front} events above 1\,GeV in the vicinity of \CB. Linear green contours of the 843\,MHz radio map are overlaid~\citep{mcadam1991}. \textcolor{\JKC}{\MOD The blue ellipse compatible with the radio lobes is also shown.}
Black cross and contours represent the estimated position and the positional errors at 68\% and 95\% confidence levels in this analysis, respectively. Magenta cross and ellipses represent the position and the errors at 68\% and 95\% confidence levels of 2FGL~J1346.6$-$6027. The position of the core of \CB\ is represented as a diamond.
\label{fig: cmap_close}}
\end{center}
\end{figure}

Figure~\ref{fig: cmap_close} shows the close-up TS map of \CB, using \emph{front} data above 1\,GeV. 
The figure clearly shows that significant discrete $\gamma$-ray emission is associated with \CB.
In addition, the figure seems to suggest that the gamma-ray emission is elongated in the direction of the radio lobes.
\textcolor{\JKC}{\MOD To test this indication quantitatively, we compare the likelihood of the spectral fit using \emph{front} and \emph{back} data in the range $0.2-200$\,GeV for a point source and an elliptical shape compatible with the extent of the radio lobes ($22'\times 10'$ in size; see Figure~\ref{fig: cmap_close}) with a uniform surface brightness. Here similarly a power-law function is assumed to model the source spectrum with its normalization and index set free during the fitting procedure (see below). 
The improvement of TS in the case of the ellipse ($ellipse1$) with respect to the point-source case, ${\rm TS}_{\rm ext}\equiv{\rm TS}_{\rm elp1}-{\rm TS}_{\rm psc}$, is 2.4. Figure~\ref{fig: cmap_close} also indicates an anisotropy of the gamma-ray emission. To test this, we rotate $ellipse1$ in $10^\circ$ steps with its center fixed and fit the data for each case. 
The TS in the case of $ellipse1$ is the highest, and TS gradually drops with the rotation to a minimum in the case of the ellipse ($ellipse2$) which is rotated by $90^\circ$ from $ellipse1$. The difference between the corresponding values of TS is ${\rm TS}_{\rm rot}\equiv{\rm TS}_{\rm elp1}-{\rm TS}_{\rm elp2}=1.4$. To test the significance of these improvements, we simulate 1000 \emph{front} and \emph{back} datasets in $0.2-200$\,GeV for the same time period as that of the observation by \textsc{gtobssim} in the Science Tools analysis package\footnotemark[1], using a model where a point source for Centaurus B, the Galactic diffuse emission, and the isotropic backgrounds are included. The point source is set at the best-fit location and has the spectral shape of a power-law function (see below). All components in the model are set at the best-fit values in the range $0.2-200$\,GeV (see \S~\ref{sec: sed}).
Each simulated dataset is fitted with models which include \CB, the Galactic diffuse emission, and the isotropic backgrounds. The normalization of all components and the index of \CB\ are set free. We compare TS between cases that the spatial shape of \CB\ is the point source, $ellipse1$, or $ellipse2$. 
Note that the position of the point source is fixed at the position estimated by using $> 1$~GeV \emph{front} data for each simulated dataset (i.e., the same estimation as that for the observed data).
The results show that the probability for obtaining ${\rm TS}_{\rm ext} \geq 2.4$ and ${\rm TS}_{\rm rot} \geq 1.4$ at once is 3\%. This indicates a weak hint of the spatial elongation along the radio lobes in the $\gamma$-ray band, although we can not significantly confirm it.}
Data that will be accumulated in the coming years of the \emph{Fermi} mission will provide us with better spatial information to discuss its extent in more detail.

We estimate the position of the $\gamma$-ray source to be ($\alpha$, $\delta$)=($206\fdg 63$, $-60\fdg 44$), following the localization of the highest TS (TS$_{\rm max} = 23$) in the TS map (Figure~\ref{fig: cmap_close}). 
\textcolor{\JKC}{\MOD The positional errors at 68\% and 95\% confidence levels are about {0\fdg 06} and {0\fdg 11}, which} are delineated by contours of decrements of TS from TS$_{\rm max}$ by amounts of 2.3 and 6.0, respectively \citep{Mattox96}. 
The estimated position with the 68\% error contour is consistent with the position of 2FGL~J1346.6$-$6027 at ($\alpha$, $\delta$)=($206\fdg 66$, $-60\fdg 47$), and contains the core of \CB\ at ($\alpha$, $\delta$)=($206\fdg 704$, $-60\fdg 408$) \citep{JLM01}. TS at the estimated position is 29 using \emph{front} and \emph{back} data in the range $0.2-200$\,GeV, corresponding to a $4.5\sigma$ detection. 
The TS value is lower than in the 2FGL catalog (TS $\sim$ 39 for $0.1-100$\,GeV), despite the longer integration time here. As we show in \S~\ref{sec: sed} the source is not significantly variable, although its measured flux was greatest in the first half of the time interval analyzed here, which overlaps most of the 2-year interval analyzed for the 2FGL catalog. In addition, the likelihood analysis for the 2FGL catalog treated \emph{front} and \emph{back} events separately, which increased the sensitivity, and TS values, slightly.
Hereafter, the $\gamma$-ray source associated with \CB\ is analyzed as a point source with this estimated position.

\subsection{Spectrum and Time Variability}
\label{sec: sed}

Figure~\ref{fig: sed} shows the spectrum of the source associated with \CB\ using \emph{front} and \emph{back} events. The spectral energy distribution (SED) is obtained by dividing the $0.2-20$\,GeV energy band into four energy bins. The indices of the other sources are fixed at the values given in the 2FGL catalog. The \CB\ source is fitted with a simple power-law function in each energy bin with the photon index fixed at 2.6, the value obtained from a broadband fitting over the larger $0.2-200$\,GeV energy range (see below). In each fit, all the sources within the ROI as well as the diffuse components in the model are fitted with their normalizations set free.

\begin{figure}[!h] 
\begin{center}
\includegraphics[width=3.5in]{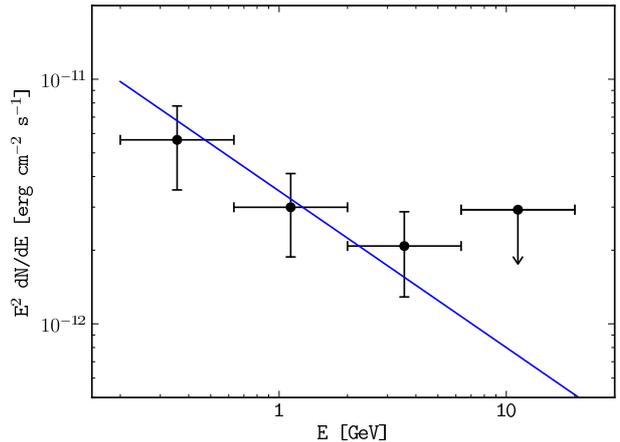}
\caption{\small \FL\ spectrum of the \CB\ source. The error bar indicates statistical errors of 1$\sigma$. The arrow represents the upper limit on the flux at 95\% confidence level. The blue line represents the best-fit power-law function from a binned likelihood fit in the $0.2-200$\,GeV range.
\label{fig: sed}}
\end{center}
\end{figure}

We tested a possible spectral steepening at the highest photon energies by performing likelihood-ratio tests between a power-law (the null hypothesis) and a a log parabola function (the alternative hypothesis) for the $0.2-200$\,GeV data. The resulting Test Statistic is ${\rm TS_{LP}} = -2\ln (L_{\rm PL}/L_{\rm LP}) = 0.0$, which means no significant curvature is present in the spectrum. The parameters obtained with the power-law model are the photon index $\Gamma_{\gamma} = 2.6 \pm 0.2$ and the integrated $0.2-200$\,GeV flux of $(1.9\pm0.5) \times 10^{-8}$\,photon\,cm$^{-2}$\,s$^{-1}$. 
\textcolor{\JKC}{\MOD We note that the uncertainties of the effective area translate to uncertainties in the flux of $\lesssim10$\% and $\lesssim0.1$ for the index~\citep{PASS7}.}
The best-fit power-law function is represented as a blue line in Figure~\ref{fig: sed}. 
We note that the flux and index obtained in the likelihood analysis change negligibly if the position of 2FGL J1346.6-6027 or the \CB\ core is used instead for the \CB\ source.

\begin{figure}[!h] 
\begin{center}
\includegraphics[width=3.5in]{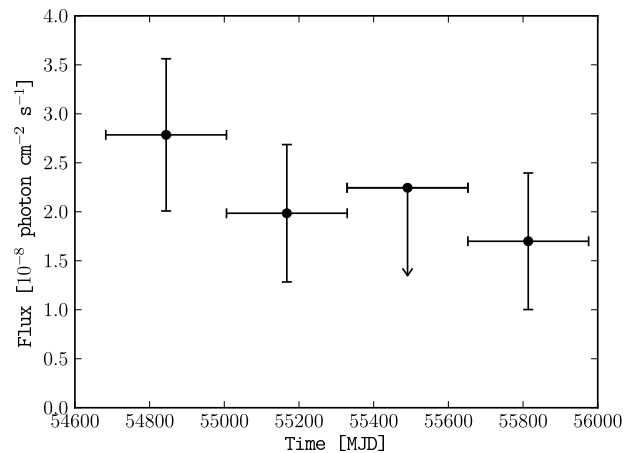}
\caption{\small \FL\ light curve of the \CB\ source in the photon energy range of $0.2-200$\,GeV. The error bar indicates statistical errors of 1$\sigma$, and the arrow represents the upper limit on the flux at 95\% confidence level.
\label{fig: lc}}
\end{center}
\end{figure}
\begin{table*}
\begin{center}
\caption{Log of \SZ\ observations}
\label{tbl:obslog}
\begin{tabular}{lcccc}
\hline \hline
Name & Observation ID & Start time (UT) & End time (UT) & Net Exposure$^{\dagger}$ (ks)\\
\hline
\CB\     & 806017010 & 2011 Jul 16 20:15:14 & 2011 Jul 18 11:50:13 & 93.6\\
4U 1344$-$60 & 705058010 & 2011 Jan 11 19:01:05 & 2011 Jan 14 11:24:11 & 93.9 \\
\hline
\end{tabular}\\
\tablefoot{
\tablefoottext{\dagger}{After the event screenings.}\\
}
\end{center}
\end{table*}

Figure~\ref{fig: lc} shows the light curve of the \CB\ source using \emph{front} and \emph{back} data in the $0.2-200$\,GeV range. We divide the whole LAT observation period (about 43 months) into only four bins, given the low flux of the source. In each time bin, the flux of the source is evaluated by performing \textsc{gtlike} with the Galactic diffuse and isotropic components fixed to the best-fit values obtained for the whole time range, and the photon index fixed at the value 2.6 obtained from a broadband fitting for the range $0.2-200$\,GeV. 
\textcolor{\JKC}{\MOD 
To test the variability quantitatively, we calculate each value of the log likelihood and sum them ($\ln(L_i)$; the alternative hypothesis). We also calculate the log likelihood obtained from the fitting for the whole period ($\ln(L_{\rm const})$; the null hypothesis). Then, ${\rm TS}_{\rm var} = -2 \ln(L_{\rm const}/L_i)$ is distributed as $\chi^2$ with three degrees of freedom~\citep{2FGL}, and we obtain $\chi^2/{\rm dof}  = 0.95$. The probability for obtaining a value $\chi^2/{\rm dof} \geq 0.95$ with three degrees of freedom is $\sim40\%$, which means that there is no indication for variability of \CB\ at $\gamma$-rays for the period spanned by the \FL\ observations.}

\section{\SZ\ data}
\label{sec: xray}

We observed \CB\ with the \SZ\ X-ray satellite \citep{Mitsuda2007} on 2011 July 16$-$18 (Observation ID 806017010). The nominal pointing position was set to the core of the radio galaxy. Since the main goal of the analysis is to search for the diffuse X-ray emission from the radio lobes, here we focus on the data taken with the X-ray Imaging Spectrometer \citep[XIS;][]{Koyama2007}, which consists of four CCD sensors each placed on the focal plane of the X-ray Telescope \citep[XRT:][]{Serlemitsos2007}. Currently two front-illuminated CCDs (XIS0 and XIS3) and one back-illuminated CCD (XIS1) are operational, and the observation was performed with all the XIS sensors set to full-window and normal clocking modes. In this paper we present the results obtained with XIS0 and XIS3 (hereafter FI). 
This is because the injection charge amount has been changed from 2\,keV to 6\,keV equivalent for the BI sensor on 2011 June 1, and the current non-X-ray background (NXB) estimation for XIS1 using \textsc{xisnxbgen} does not reproduce accurately the actual level of NXB\footnote{http://heasarc.gsfc.nasa.gov/docs/suzaku/analysis/abc/node8.html}.
In the spectral analysis, we also ignore the data between 1.8 and 2.0 keV, because there exists a calibration uncertainty around the Si-K edge\footnote{http://heasarc.gsfc.nasa.gov/docs/suzaku/analysis/abc/node8.html}.

Since the radio lobes of \CB\ are extended over almost the whole \SZ\ CCD chip, it is difficult to extract background information from the data. Hence, we also analyze the \SZ\ data for 4U 1344-60 (Observation ID 705058010; see also \cite{4U1344}), which is located $\sim 14'$ away from \CB. 
{\MOD We assume that the diffuse X-ray backgrounds are the same between these data given the proximity.}
This observation was performed on 2011 January 11$-$14 with full-window and normal clocking modes.

The screening criteria of XIS data are as follows. We utilized events with GRADE 0, 2, 3 ,4 and 6, then removed flickering pixels by using \textsc{cleansis}. Good-time intervals were selected by the same criteria as described in {\it The Suzaku Data Reduction Guide}\footnote{http://heasarc.nasa.gov/docs/suzaku/analysis/abc/}, namely ``SAA\_HXD$==$0 $\&\&$ T\_SAA\_HXD$>$436 $\&\&$ ELV$>$5 $\&\&$ DYE\_ELV$>$20 $\&\&$ ANG\_DIST$<$1.5 $\&\&$ S0\_DTRATE$<$3 $\&\&$ AOCU\_HK\_CNT3\_NML\_P$==$1''. In the screening and the following analysis, we used the \textsc{HESoft}\footnote{http://heasarc.nasa.gov/lheasoft/} version 6.11 and calibration database (\textsc{CALDB})\footnote{http://heasarc.gsfc.nasa.gov/docs/heasarc/caldb/} released on 2012 February 11. The observation log and net exposures after the event screenings are summarized in Table 1.

\subsection{Core}

Figure~\ref{fig:xisimage} shows the FI image in the $2-10$\,keV band after subtracting the NXB and correcting for vignetting and exposure.
We extracted the events in a circular region with a radius of $4'$ centered at the \CB\ core position and produced the XIS spectrum shown in Figure~\ref{fig:core_spe}. 
The background region was selected from blank sky within the field of view (see Figure~\ref{fig:xisimage}). 
Response and ancillary response files were generated by \textsc{xisrmfgen} and \textsc{xissimarfgen}, respectively.
Then, we fitted the $0.6-9.0$\,keV data with an absorbed power-law model (\textsc{wabs*powerlaw}) in \textsc{XSPEC}.
The best-fit parameters are summarized in Table~\ref{tbl:core_spe}.

\begin{figure}[!h] 
\begin{center}
\includegraphics[width=3.5in]{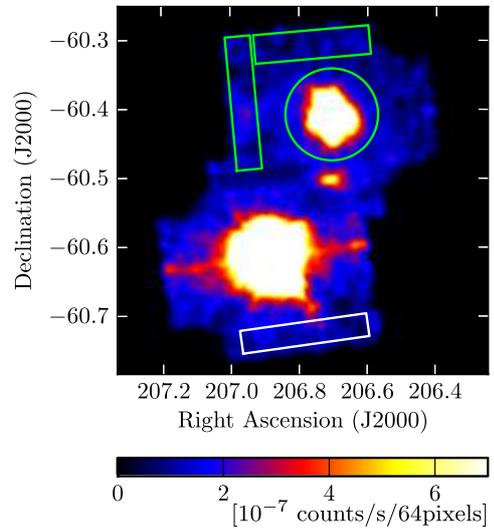}
\caption{\small \SZ\ FI image ($2-10$\,keV) of \CB. The upper bright point-like source is the nucleus of the \CB, and the lower source is 4U 1344$-$60. 
Green circle denotes the source extracted region for the \CB\ core. 
Two green rectangles on the same CCD indicate the background regions. 
The diffuse X-ray background spectrum was produced from events within the lower white rectangular region.
\label{fig:xisimage}}
\end{center}
\end{figure}
\begin{table}
\begin{center}
\caption{Best-fit model parameters for the core of \CB\ and the source N}
\label{tbl:core_spe}
\begin{tabular}{lcc}
\hline \hline
Parameter & Core & Source N\\
\hline
$N_{\rm H}\,[\times 10^{22}\,{\rm cm}^{-2}]$ & $1.63\pm0.08 $ & $1.55^{+0.56}_{-0.46}$ \\
$\Gamma_{\rm X}$   & $1.62\pm0.05$ & 2.02$^{+0.35}_{-0.32}$ \\
Normalization & $\left( 1.10^{+0.09}_{-0.08} \right) \times 10^{-3}$ & $\left( 1.02^{+0.72}_{-0.39} \right) \times 10^{-4}$\\
\hline
Unabsorbed flux & $\left( 5.1\pm 0.4 \right) \times 10^{-12}$  & $2.57 \times 10^{-13}$\\
~~~$[$erg\,cm$^{-2}$\,s$^{-1}$$]$  & & \\
\hline
$\chi^2_{\nu} \, (\chi^2/d.o.f)$  & 1.18 (366.13/310) & 1.30 (23.37/18)\\
\hline
\end{tabular}\\
\tablefoot{
\tablefoottext{\dagger}{$N_{\rm H}$ and $\Gamma_{\rm X}$ are absorption column density and photon index, respectively.
The unabsorbed flux were integrated in 2--10\,keV range. Errors are 90\% confidence level.}\\
}
\end{center}
\end{table}

\begin{figure}[!h] 
\begin{center}
\includegraphics[angle=-90, width=3.5in]{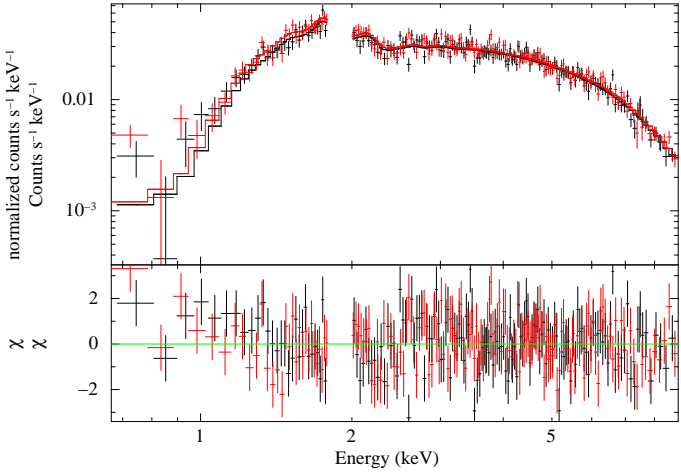}
\caption{\small \SZ/XIS spectrum of the \CB\ core. XIS0 and XIS3 spectra are shown in black and red, respectively.
\label{fig:core_spe}}
\end{center}
\end{figure}

\subsection{Diffuse Background}

A spectral analysis of faint diffuse emission requires the precise evaluation of the X-ray background emission.
We extracted the background spectrum using the blank field of 4U 1344$-$60 (the $11'\times 2'$ rectangular region shown in Figure~\ref{fig:xisimage}).
Here we consider contamination from 4U 1344$-$60 given a relatively large PSF ($\sim2'$) of \SZ\ XRT.
FI count rate of the contamination simulated by \textsc{xissim}~\citep{XISSIM} was $0.0023$\,cts\,s$^{-1}$ in the $0.5-7.0$\,keV range, while the measured count rate was $0.0114$\,cts\,s$^{-1}$.
We thus found that about 20\% of the X-ray events within the considered rectangle are contaminated by 4U 1344$-$60.
Note that we have analyzed 14 ks Swift/XRT archival data and confirmed that there are no point sources in the rectangular region.
{\MOD We derive the 95\% confidence level upper limit for the unabsorbed flux of $1.1 \times 10^{-13}~{\rm erg\ cm^{-2}\ s^{-1}}$ at $0.5-10$~keV, assuming a single power law with index 2.0 and $N_{\rm H} =1.05 \times 10^{22}~{\rm cm^{-2}}$.}

The diffuse X-ray background consists of three components: thermal emission from the Local Hot Bubble, a thermal component due to the Galactic halo, and a power-law component originating from a superposition of unresolved AGNs (``Cosmic X-ray Background'', CXB). 
After subtracting both the NXB estimated with \textsc{xisnxbgen} and the contaminant of 4U 1344$-$60, 
we fitted the background data in the $0.6-8.0$\,keV range with a model of \textsc{mekal+wabs*(mekal+powerlaw)}~\citep[see e.g.,][]{Yuasa09}.
In the fit, the absorption column density was fixed to the value in that direction\footnote{http://heasarc.nasa.gov/cgi-bin/Tools/w3nh/w3nh.pl} ($N_{\rm H} = 1.06 \times 10^{22}$\,cm$^{-2}$).
The value of $N_{\rm H}$ obtained in the spectral analysis of the core is larger than this value possibly due to the absorption of the host galaxy.
The best-fit spectra and parameters are shown in Figure~\ref{fig:bg_spe} and Table~\ref{tbl:xrb_parameters}, respectively. 
The derived power-law index does not agree with the ``standard'' CXB index of 1.41, but is similar to that of the Galactic ridge emission (index $\sim 2.1$, see e.g., \citet{Revnivtsev2006}). 
This could be explained by the fact that the analyzed field is located close to the Galactic plane.

\begin{table}
\begin{center}
\caption{Best-fit parameters for the diffuse X-ray background spectrum derived from blank sky around 4U 1344$-$60}
\label{tbl:xrb_parameters}
\begin{tabular}{llc}
\hline \hline
Component & Parameter & Value\\
\hline
Local Hot Bubble & $kT\,[{\rm keV}]$ & 0.61$^{+0.09}_{-0.16}$\\
                & $Z$ & 1.0 (fixed) \\
                & Normalization & $\left( 7.0 \pm 3.3 \right) \times 10^{-4}$\\
                \hline
  & $N_{\rm H}$\,[$\times 10^{22}$\,cm$^{-2}$] & 1.06 (fixed)\\
Galactic halo    & $kT$\,(keV)  & 0.17$^{+0.09}_{-0.06}$ \\
                & $Z$  & 1.0 (fixed) \\
                & Normalization  & 0.26$^{+3.38}_{-0.24}$ \\
CXB & $\Gamma_{\rm X}$  & 2.22$^{+0.23}_{-0.22}$ \\
                & Normalization & $\left( 6.6^{+1.8}_{-1.5} \right) \times 10^{-3}$ \\
                \hline
                & $\chi^2_{\nu} (\chi^2/d.o.f)$  & 1.02 (63.41/62) \\
                \hline
\end{tabular}
\tablefoot{
\tablefoottext{\dagger}{$kT$ and $Z$ are temperature and metal abundance in terms of the solar value of the plasma, respectively.
Errors are 90\% confidence level.}\\
}
\end{center}
\end{table}

\begin{figure}[!h] 
\begin{center}
\includegraphics[angle=-90, width=3.5in]{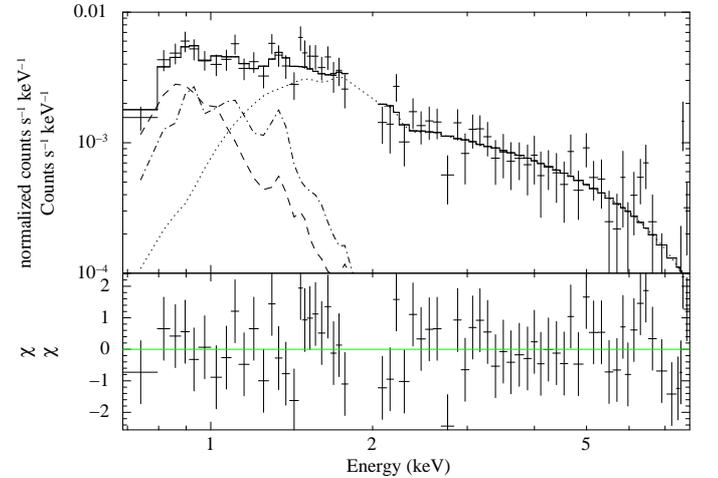}
\caption{\small \SZ/XIS FI spectrum of the X-ray background obtained from blank sky near 4U 1344$-$60. 
Dashed, dot-dashed, and dotted lines represent the components of the Local Hot Bubble, the Galactic halo, and the CXB, respectively.
\label{fig:bg_spe}}
\end{center}
\end{figure}

\subsection{Lobes}

\begin{figure}[!h] 
\begin{center}
\includegraphics[width=3.9in]{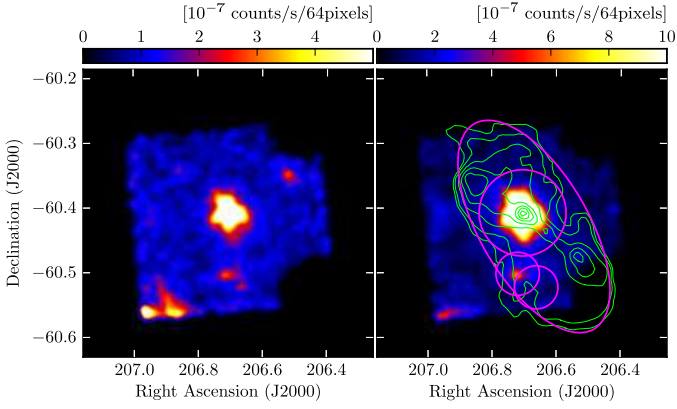}
\caption{\small (Left) \SZ\ FI $0.5-2$\,keV image of \CB.
(Right) \SZ\ FI $2-10$\,keV image of \CB. The central bright source is the nucleus of the system. Green contours denote the 843\,MHz radio intensity obtained from \citet{mcadam1991}. Magenta ellipse in the image denotes the source extracted region for the lobes. The three magenta circular regions containing X-ray point sources are excluded from the analysis of the emission from the lobes.
\label{fig:lobe}}
\end{center}
\end{figure}

We found two faint point sources within the {\MOD radio} contours of the lobes of \CB\ in the $0.5-2$~keV and $2-10$~keV Suzaku FI NXB-subtracted exposure-corrected images (see Figure~\ref{fig:lobe}). 
After masking the core and these two X-ray sources with $4'$ and $2'$ radii circles,
the elliptical region ($22'\times10'$ in size) was extracted for the spectral analysis of the lobes.}
NXB was estimated with \textsc{xisnxbgen}. The contamination from the \CB\ core was simulated by \textsc{xissim}, where the values in Table~\ref{tbl:core_spe} were used for the spectra of the core. The spectrum of diffuse X-ray background of this region was {\MOD simulated} with \textsc{XSPEC} by using the values tabulated in Table~\ref{tbl:xrb_parameters}. 
Figure~\ref{fig:plot_all} shows the observed spectrum together with the three background components.
We utilized \textsc{xissimarfgen} to produce the ancillary response file assuming {\MOD a uniform surface brightness in the elliptical region.}

\begin{figure}[!h] 
\begin{center}
\includegraphics[angle=-90, width=3.2in]{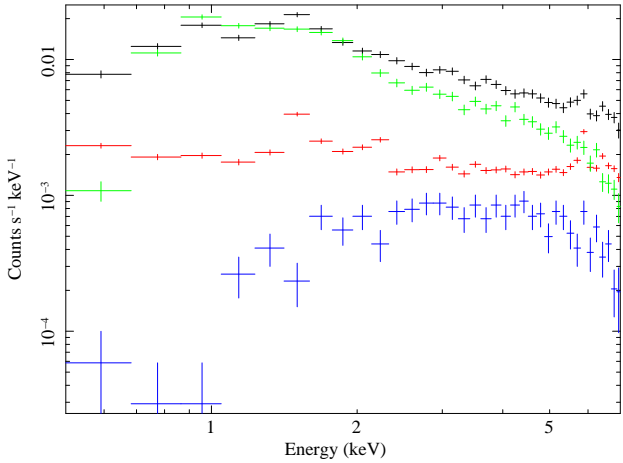}
\caption{\small \SZ/XIS FI spectrum of the \CB\ lobe region. The observed spectrum are shown in black. 
Green, red and blue spectra denote the background components: the diffuse X-ray background, the NXB, and the contaminant from the core, respectively.
\label{fig:plot_all}}
\end{center}
\end{figure}

Since the X-ray events of the diffuse background emission within the selected background region are small, statistical errors of the derived parameters are relatively large (see Table~\ref{tbl:xrb_parameters}). 
In particular, the error in the normalization of the power-law component of the X-ray background {\MOD significantly} affects the {\MOD evaluation} of the faint diffuse X-ray emission from the radio lobes. As shown in Table 3, the $\sim$25\% statistical error in normalization dominates the $\sim$3\% systematic error in the estimation of NXB spectrum~\citep{Tawa08} up to about 6 keV (see Figure~\ref{fig:bg_spe}). 
Hence, we increased/decreased the normalization of power-law component by 25\% and produced faked X-ray background spectrum. Then after subtracting it we searched for the residual emission above 2 keV. 
When we increased the normalization by 25\%, the residual spectrum does not show significant power-law component and the best-fit normalization corresponded to zero ($\Gamma_{\rm X}=1.7$ is assumed following the radio data; see \S~\ref{sec: radio}). To derive the upper limit for the lobe X-ray emission, we considered the most conservative case, namely the normalization of the power-law component of X-ray background decreased by 25\%. Then we fitted the residual spectrum by assuming an absorbed power-law shape with only normalization set free and the other parameters fixed as $\Gamma_{\rm X}=1.7$ and $N_{\rm H}=1.06 \times 10^{22}$\,cm$^{-2}$.\footnotemark[9] 
The obtained 90\% confidence level upper limit for the unabsorbed $2-10$~keV emission of the \CB\ lobes is $<8.7 \times 10^{-13}$\,erg\,cm$^{-2}$\,s$^{-1}$.

\begin{figure}[!h] 
\begin{center}
\includegraphics[angle=-90, width=3.5in]{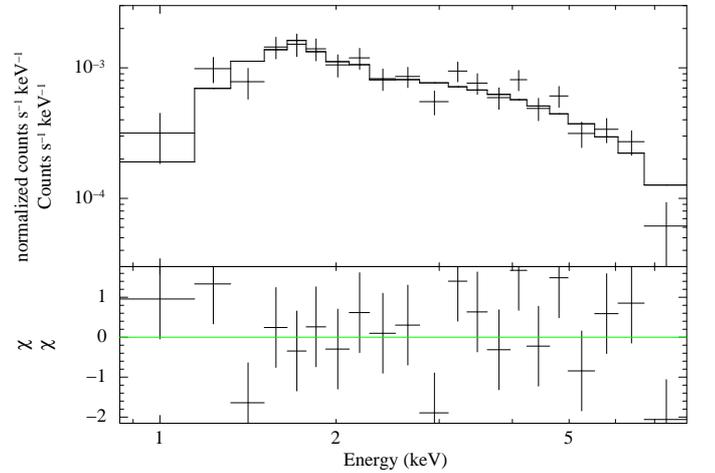}
\caption{\small \SZ/XIS FI spectrum of the source N.
\label{fig:hxray_spe}}
\end{center}
\end{figure}

As shown in Figure~\ref{fig:lobe}, one of the masked sources (source~N), located at ($\alpha$, $\delta$)=($206\fdg 721$, $-60\fdg 506$), has a hard X-ray spectrum.
This source is positionally coincident with the diffuse-like emission seen in the \emph{ASCA} GIS $1.5-3.0$\,keV image within the South-Western radio lobe \citep[see Figure~3 in][]{tashiro98}. 
To derive spectral parameters of the source~N, we extracted a circular region with a radius of $1'$ centered at this source, and produced a FI spectrum as shown in Figure~\ref{fig:hxray_spe}. 
The background region was selected from the blank field of the same CCD chip. 
The spectrum was well represented by the absorbed power-law model as shown in Table~\ref{tbl:core_spe}.
We also searched for possible counterparts of source N in other wavelength using SIMBAD\footnote{http://simbad.u-strasbg.fr/simbad/}, and found an infrared source just at the position of source N. This coincidence again supports the possibility that the point-like X-ray emission is unrelated to the radio lobe.

\section{Radio data}
\label{sec: radio}

An 843\,MHz image \citep{mcadam1991} from the Molonglo Observatory Synthesis Telescope (MOST), with $43''$ resolution and excellent surface brightness sensitivity, provides a reliable total integrated flux density of $\sim 150$\,Jy. The northeastern lobe is substantially brighter with an integrated flux of $\sim 93$\,Jy while the southwestern lobe contains $\sim 57$\,Jy. \citet{JLM01} used integrated flux densities from the literature, ranging from 30\,MHz to 5\,GHz, to derive a radio spectral index of $\alpha=0.73$. As noted in \S~\ref{sec: intro}, the 1.4\,GHz radio power puts \CB\ near the boundary of the FR\,I/FR\,II luminosity division. The bright jet and diffuse lobe emission resembles the morphology of an FR\,I radio source, while the northeastern lobe has a sharp edge reminiscent of an FR\,II radio galaxy~\citep[see in this context][]{GK00,Laing11}. 

\cite{JLM01} also present observations of the entire source using the Australian Telescope Compact Array (ATCA) at 4.8 and 8.6\,GHz. We are unable to obtain reliable integrated flux densities for the lobes from these images because of the large angular extent of the source ($\simeq 24'$). A substantial amount of flux is lost, since the shortest baseline between antennas ($\sim 20$\,m for mosaicked images from the ATCA) is too large to recover the emission on the largest spatial scales. The entire source has also been imaged at 20\,GHz \citep{burke2009}. Again, the integrated fluxes cannot be used in our SED modeling due to the missing short spacings, although the results provide an accurate representation of the core, jet and hotspot structures. 
Also, in this particular case, potential imaging artifacts due to the bright core may contaminate the faint lobe emission at 20\,GHz. 

Since our line of sight to the source passes within $1.7^{\circ}$ of the Galactic plane, single dish measurements can be contaminated by the Galactic emission (particularly at low radio frequencies). 
The Parkes-MIT-NRAO survey conducted at 4.8\,GHz \citep{pmn1993} provides us with a second suitable image for extraction of integrated fluxes. The Parkes 64\,m dish image has high enough angular resolution ($\sim 5'$) to resolve the source into two lobes, with integrated flux densities of $\simeq 29$\,Jy and $\simeq 18$\,Jy for the northeastern and southwestern lobes, respectively. 
Furthermore, the observing frequency is also high enough, at this spatial resolution, to avoid any substantial contamination from diffuse Galactic emission. The core and inner jet flux contributes $\sim$3~Jy at 5~GHz \citep{marshall05}, which is a small fraction of the total integrated emission of $\sim$47~Jy. Since the lobes are brighter at lower frequencies, the contribution from the core becomes negligible at frequencies below 5~GHz. Hence, the integrated fluxes from 30~MHz to 5~GHz provide accurate representations of the integrated emission from the lobes.

The newly launched {\em Planck} satellite \citep{Planck} has observed the sky
in nine frequency bands from 30--857~GHz with the first all-sky
compact-source catalogue of high-reliability sources presented in \cite{PlanckERCSC}.
They used detailed source extraction algorithms to reliably detect discrete
sources immersed in a diffuse foreground (e.g., the Galactic plane).
From their catalogue, we were able to extract integrated fluxes for Centaurus~B
at 30, 44, 70, 100 and 143~GHz. The spatial resolution of $33'$ at 30~GHz
means that Centaurus~B is unresolved and the total flux density accurately
represents the entire flux (lobes plus core) from the source at this frequency.
We cannot discriminate between the contribution of the core and lobe fluxes
from the available data at higher radio frequencies, however, we expect the flat-spectrum
core to become increasingly dominant over the steep-spectrum lobes above 30~GHz.
The extracted 44~GHz flux density appears to slightly underestimate the total flux of
the source at this frequency (based on power-law fit to all {\em Planck} data points),
even though the spatial resolution of $27'$ should be sufficient to include
the entire emission from the source. The spatial resolution of {\em Planck} at 70, 100
and 143~GHz of $13'$, $10'$ and $7'$ means that a large fraction of the flux from
the lobes of Centaurus~B is missing. 

\section{Discussion}
\label{sec: discussion}

The observed GeV emission of \CB\ could be produced within the unresolved core of the system, in the analogy to some other radio galaxies detected by the \FL\ such as NGC\,1275 \citep{PerA,kataoka10}, M\,87 \citep{M87,M87MWL}, 3C\,120 or 3C\,111 \citep{BLRG}. 
In those cases, the broad-band nuclear spectra as well as the established $\gamma$-ray variabilities indicate that the $\gamma$-ray production is dominated by the inner (pc/sub-pc scale) relativistic jets observed at intermediate viewing angles (``misaligned blazars''). 
The jet/counterjet radio brightness and polarization asymmetries on slightly larger scales in the source studied here imply intermediate jet inclinations and relativistic beaming involved, supporting the idea that the nucleus may contribute to the observed $\gamma$-ray emission, at least at some level.
On the other hand, the LAT analysis of \CB\ provides a weak hint of the spatial extension of the $\gamma$-ray counterpart, as well as no indication of significant flux variability in the GeV range, which suggest that the extended (100-kpc scale) lobes in the system could be considered as a promising emission site of the high-energy photons as well. The arguments based on the morphology of the analyzed $\gamma$-ray emitter and its timing properties are nonetheless rather weak, due to the limited photon statistics.

\begin{figure}[!h] 
\begin{center}
\includegraphics[width=3.5in]{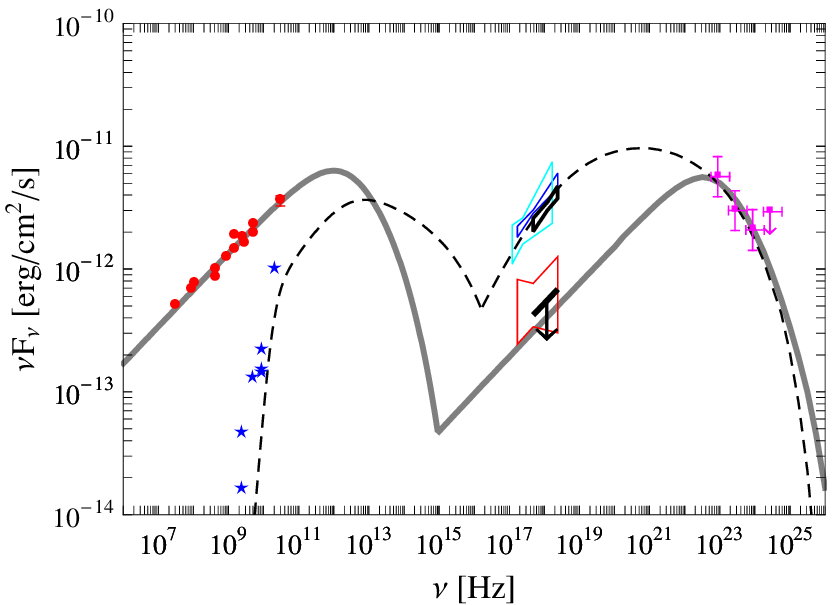}
\caption{\small Broad-band SEDs of \CB\ and the model curves for the emission of the extended lobes and core in the system.
Red circles denote the total radio fluxes within the range $30$\,MHz\,$-5$\,GHz~\citep[and references therein]{JLM01} and at $30$\,GHz measured by \emph{Planck}.
The radio fluxes for the unresolved core, plotted as blue stars, are taken from \citet{JLM01}, \citet{Fey04} and \citet{burke2009}. The \emph{ASCA} measurements of the lobes' and core spectra within the range $0.7-10$\,keV, as discussed in \citet{tashiro98}, are denoted by red and blue bow-ties, respectively. The \emph{Chandra} $0.5-7$\,keV spectrum of the unresolved core is given by cyan bow-tie following \citet{marshall05}. The \FL\ fluxes represented by magenta squares correspond to the analysis presented in \S~\ref{sec: obs} of this paper. The black bow-tie denotes the $2-10$\,keV emission of the nucleus emerging from the analysis of our \SZ\ data, while the thick black line shows the $2-10$\,keV \SZ\ upper limit foe the lobes' emission (see \S~\ref{sec: xray}). Thick gray curves denote the broad-band emission of the lobes modeled in in the framework of the synchrotron and inverse-Compton scenario, while black dashed curve corresponds to the SSC model of the unresolved core in the system, as discussed in \S~\ref{sec: discussion}.
\label{fig: model}}
\end{center}
\end{figure}

At X-ray frequencies, the core of \CB\ appears as a relatively bright source. In their analysis of the \emph{ASCA} data, \citet{tashiro98} found the $2-10$\,keV nuclear unabsorbed flux of $\simeq (6.4 \pm 0.2) \times 10^{-12}$\,erg\,cm$^{-2}$\,s$^{-1}$ with a power-law model of $\Gamma_{\rm X} \simeq 1.64 \pm 0.07$ and $N_{\rm H} \simeq (1.76 \pm 0.11) \times 10^{22}$\,cm$^{-2}$ possibly in excess over the Galactic value. This is in agreement with the \emph{Chandra} snapshot obtained by \citet{marshall05}, who measured the $1$\,keV nuclear flux $870^{+240}_{-180}$\,nJy and the best-fit power-law model with $\Gamma_{\rm X} \simeq 1.63 \pm 0.17$. 
Our new \SZ\ exposure confirms the previous studies, revealing the unabsorbed $2-10$\,keV flux of the core $\simeq (5.1 \pm 0.4) \times 10^{-12}$\,erg\,cm$^{-2}$\,s$^{-1}$ marginally lower than that implied by the \emph{ASCA} observations {\MOD but consistent with that detected by XMM-\emph{Newton} \citep{tashiro05}}, with $\Gamma_{\rm X} \simeq 1.62 \pm 0.05$ and $N_{\rm H} \simeq ( 1.63 \pm 0.08 ) \times 10^{22}$\,cm$^{-2}$. The unabsorbed X-ray ($2-10$\,keV) luminosity of the core, $L_{\rm X} \simeq 2 \times 10^{42}$\,erg\,s$^{-1}$, is roughly comparable to the $\gamma$-ray ($0.2-200$\,GeV) luminosity of the \CB\ system, $L_{\gamma} \simeq 6 \times 10^{42}$\,erg\,s$^{-1}$ (see Figure~\ref{fig: model}). 
{\MOD In addition, the inner parts of the main (SW) jet were resolved at arcsec scale with \emph{Chandra}~\citep{marshall05}. 
The detection of the SW jet on larger scales was also claimed in the preliminary analysis of the XMM data~\citep{tashiro05}.}
\emph{If} the observed X-ray continuum of the nucleus corresponds to the \emph{beamed non-thermal} (inverse-Compton) emission of the unresolved jet, the observed $\gamma$-ray flux could be explained well as a high-energy tail of the same nuclear emission component.

A caveat is that in radio galaxies the observed X-ray nuclear emission is typically dominated by the accretion disks and disk coronae, and not by the misaligned unresolved jets. This seems to be particularly robust in the case of powerful objects accreting at high rates, like Broad-Line Radio Galaxies for example \citep[e.g.,][]{eracleous00,BLRG}. 
For the low-power objects accreting at low rates, such as FR\,I radio galaxies, their entire observed X-ray continua may be due to the innermost parts of the jets \citep{DD04,evans06,HEC09}. This controversy relates in particular to low-power radio galaxies detected by the \FL\ \citep[see the discussion in][and references therein]{PerA,M87,CenACore}. 
\CB\ seems to be an intermediate case between low- and high-power sources, and hence one may expect a mixture of the nuclear jet and accretion-related emission components at X-ray frequencies; the observed spectral properties of the X-ray core in the system suggest however a dominant contribution from the jet.

In order to test the hypothesis that the observed $\gamma$-ray emission of \CB\ originates in the innermost parts of a misaligned relativistic jet, we apply the synchrotron self-Compton (SSC) model to the collected broad-band data of the unresolved core in the system\footnote{Note that as the observed core radio fluxes correspond to a superposition of different self-absorbed jet components, in the fitting procedure we consider these as upper limits for the emission of the modeled emission site, for which the rest-frame turnover frequency related to the synchrotron self-absorption (SSA) process reads as $\nu'_{\rm ssa} = 9$\,GHz.}, assuming that the emission site can be approximated as a homogeneous spherical region with radius $R'$, filled uniformly with tangled magnetic field and radiating electrons.
This is the first time that the SSC model is applied to the misaligned relativistic jet in \CB\ and compared with the compiled multiwavelength dataset for the unresolved core.
As shown in Figure~\ref{fig: model}, the model successfully explains the data for the following parameters: jet radius $R'=0.03$\,pc, jet magnetic field $B = 0.095$\,G, and jet Doppler factor $\delta = 2$. For the radiating electrons we assume a power-law energy spectrum $dN'_e(\gamma)/d\gamma \propto \gamma^{-2.2}$, where $\gamma$ is the Lorentz factor, between energies $\gamma_{\rm min} = 30$ and $\gamma_{\rm br} = 2 \times 10^3$, breaking to $dN'_e(\gamma)/d\gamma \propto \gamma^{-3.2}$ above $\gamma_{\rm br}$ and continuing as such up to $\gamma_{\rm max} = 10^5$. The fit requires normalization of the energy spectrum such that the ratio of the comoving electron and magnetic energy densities is $\eta'_{\rm eq} \equiv U'_e / U'_{\rm B} = 100$. All these parameters are broadly consistent with the ones inferred for the cores of other GeV-detected radio galaxies \citep[see][]{PerA,M87,CenACore}. Also the jet power emerging from the model presented here should be considered as reasonable: assuming no significant jet radial structure, the model value of the jet Doppler factor $\delta = 2$ implies jet bulk Lorentz factors $\Gamma_j = 4-7$ for the jet viewing angles within the range $\theta_j = 20^\circ-25^\circ$, and 
hence the jet luminosity, roughly, $L_j \simeq \pi {R'}^2 c \Gamma_j^2 U'_e \sim 10^{45}$\,erg\,s$^{-1}$. This estimate neglects likely contribution of protons to the jet kinetic flux, and therefore should be considered as a lower limit only.
{\MOD The implied jet-counterjet brightness asymmetry, $R \equiv [(1+\beta_j \cos \theta_j )/ (1-\beta_j \cos \theta_j)]^{\varepsilon} = (2 \Gamma_j \delta - 1)^{\varepsilon}$, where $\beta_j = \sqrt{1-\Gamma_j^{-2}}$ is the jet bulk velocity and $\varepsilon = 2+\alpha$ assuming a steady jet and the emission spectral index $\alpha$, is on the other hand rather high, $R \gtrsim 10^3$ for $\alpha \gtrsim 0.5$, much greater than the one measured on larger scales \citep{JLM01}. This is however consistent with the established idea of the jet deceleration from sub-pc to kpc scales --- the process inevitably accompanied by the formation of radial stratification of the jet --- in low and intermediate power radio galaxies (FR\,Is and FR\,I/FR\,IIs) such as \CB\ \citep[see in this context, e.g.,][]{laing99,laing02}.}

Let us now focus on the extended lobes in the \CB\ system. For modeling the broad-band SED of the lobes we use the the total radio fluxes in the range $30$\,MHz\,$-5$\,GHz as provided by \citet{JLM01}, as well as the 30\,GHz flux measured by \emph{Planck}, which should represent the integrated emission of the lobes (see \S~\ref{sec: radio}). We also take into account the X-ray upper limit of our \SZ\ observations, $F_{\rm 2-10\,keV} < 8.7 \times 10^{-13}$\,erg\,cm$^{-2}$\,s$^{-1}$ assuming $\Gamma_{\rm X}=1.7$ as shown in \S~\ref{sec: xray}. We note that this upper limit is only marginally consistent with the flux found by \citet{tashiro98}, $F_{\rm 2-10\,keV} \simeq (9.8 \pm 3.8) \times 10^{-13}$\,erg\,cm$^{-2}$\,s$^{-1}$. 
We assume that the radio emission is due to the synchrotron radiation of non-thermal electrons distributed isotropically within the homogeneous lobes. We approximate the lobes for simplicity as two spheres with radii $R \simeq 100$\,kpc, so that the total volume of the emission region is $V_{\ell} \simeq 2 \times (4\pi/3) R^3$. 
The high-energy continuum, on the other hand, including the X-ray and $\gamma$-ray fluxes, is ascribed to the inverse-Compton (IC) emission of the same electron population.
As in the case of Centaurus\,A~\citep{CenALobes}, we consider three relevant target photon fields:
the CMB, the extragalactic background light (EBL) photons within the infrared--to--optical range, and the starlight of the host galaxy. 
In our calculations we use the EBL model of \citet{RM08}, and note that this particular choice does not affect the obtained results. 
For the starlight of the host galaxy, we use a template for the giant elliptical normalized to the $1.4 \times 10^{14}$\,Hz flux of extinction-corrected flux of 0.42\,Jy \citep{laustsen77,schroeder07}. 
This gives a V-band luminosity for the host of $L_{\rm V} \simeq 10^{44}$\,erg\,s$^{-1}$, and the corresponding energy density of the starlight within the lobes $U_{\rm star} \simeq L_{\rm star} / 4 \pi R^2 c \sim 3 L_{\rm V} / 4 \pi R^2 c \lesssim 10^{-14}$\,erg\,cm$^{-3}$. The results of the modeling are presented in Figure~\ref{fig: model}.

As shown in Figure~\ref{fig: model}, the simplified model \emph{can} explain the radio and $\gamma$-ray fluxes well, \emph{not violating the X-ray upper limit}.
For the model parameters, the IC spectrum is dominated by the CMB component, with negligible contribution from the EBL and the starlight of the host galaxy.
A broken power-law electron energy distribution $dN_e/d\gamma$ is set to $\propto \gamma^{-2.4}$ for $\gamma < \gamma_{\rm br}$ and $dN_e/d\gamma \propto \gamma^{-3.4}$ for $\gamma > \gamma_{\rm br}$, with minimum, break, and maximum Lorentz factors of $\gamma_{\rm min} = 1$, $\gamma_{\rm br} = 3.5 \times 10^5$, and $\gamma_{\rm max} = 1.5 \times 10^6$, respectively. 
The lobes appear to be slightly out of the energy equipartition, with the corresponding electron--to--magnetic field energy density ratio $\eta_{\rm eq} \equiv U_e / U_{\rm B} \simeq 4$, and the magnetic {\MOD field intensity $B \simeq 3.5$\,$\mu$G \citep[see in this context][]{tashiro98,kataoka05,croston05,isobe11}. }
The lobe pressure stored in ultrarelativistic electrons and magnetic field is $p_{\rm e+B} \simeq 8 \times 10^{-13}$\,dyn\,cm$^{-2}$, while the total energy injected by the jets over the source lifetime (a sum of the work done in displacing a volume $V_{\ell}$ of surrounding gas at pressure $p_{\rm e+B}$ and the energy of the material inside the cavity) is $E_{\rm tot} \simeq 4 \, p_{\rm e+B} \, V_{\ell} \sim 8 \times 10^{59}$\,erg. 
These values, derived from the crude approximations in the modeling, are typical for intermediate-power {\MOD radio galaxies~\citep[see e.g., ][]{croston05,Birzan08,isobe09}. }

It can be seen from Figure~\ref{fig: model} that the model fit depends crucially on the high-energy tail of the electron energy distribution within the lobes, which is currently unconstrained by the data. To account for the \FL\ fluxes, the electron break Lorentz factors within the lobes must be as high as $\gamma_{\rm br} = 3.5 \times 10^5$, giving the expected break synchrotron frequency $\nu_{\rm br} \simeq 1.8 \times 10^{12}$\,Hz for $B \simeq 3.5$\,$\mu$G. This would then correspond to the spectral age of the lobes
\begin{equation}
\tau_{\rm rad} \simeq 50.3 \, \frac{(B/10\,\mu {\rm G})^{1/2}}{(B/10\,\mu {\rm G})^{2} + 0.1} \, \left(\frac{\nu_{\rm br}}{\rm GHz}\right)^{-1/2} \, {\rm Myr} \sim 3 \, {\rm Myr},
\end{equation}
\noindent
according to the standard synchrotron aging theory \citep[see, e.g.,][]{konar06,machalski07}. This is a rather young age for the observed $\sim 400$\,kpc extension of the entire radio structure, implying either particularly high advance velocity of the jets in the system, $v_{\rm adv} \gtrsim 0.1 \, c$ in order for the dynamical age $\tau_{\rm dyn} \sim 2 R/v_{\rm adv}$ to be comparable with $\tau_{\rm rad}$, or efficient re-acceleration of ultrarelativistic particles stored within the lobes. 
Such a re-acceleration process (the exact nature of which we do not specify) would result in a younger spectral appearance of the lobes. In this case, the age derived based on the spectral analysis, $\tau_{\rm rad} \sim 3$\,Myr, could be much shorter than the true age of the system, namely $\tau_{\rm dyn} \sim 2 R/v_{\rm adv} > 10$\,Myr {\MOD with the jet advance velocities $v_{\rm adv} < 0.1 \, c$, which are typically derived for the evolved radio galaxies \citep[linear sizes $>10$\,kpc; see the discussions in][]{scheuer95,alexander00,machalski07,kawakatu08}.}

Here we favor the re-acceleration scenario, since the alternative explanation involves uncomfortably high values of $v_{\rm adv} \sim 0.1 \, c$, exceeding by orders of magnitude the sound velocity in the surrounding medium $c_s = \sqrt{5 k T / 3 m_p} \sim 0.001 \, (T/10^7\,{\rm K})^{1/2} \, c$ (with the expected $T\sim 10^7$\,K). {\MOD This would imply the presence of a strong bow-shock driven in the ambient medium by the supersonically expanding lobes, with the Mach number $\mathcal{M} =v_{\rm adv} / c_s \gg 10$ unprecedentedly high as for an evolved radio galaxy; we emphasize that all the bow-shocks discovered till now in such systems are characterized by very low Mach numbers, typically $\mathcal{M} \sim 2$ \citep[see][and references therein]{Siemiginowska12}, with the prominent exception of the  $\mathcal{M} \simeq 8$ shock detected around the Southern inner lobe in the Centaurus\,A system, albeit on much smaller (kpc) scales \citep{Croston09}.} We also note that similar conclusions regarding particle re-acceleration in the extended lobes have been presented in the case of two other radio galaxies resolved by the \FL, Centaurus\,A \citep{CenALobes} and NGC\,6251 \citep{takeuchi12}. 
Clearly, high-frequency ($> 30$\,GHz) radio observations by ground-based interferometers with the appropriate resolution and sufficiently short baselines are required to make further robust statements in this respect, since the association of the $\gamma$-ray emission site in \CB\ could be confirmed further by constraining the high-energy tail of the lobes' synchrotron continuum.

It is also important to determine the level of the diffuse non-thermal X-ray emission in \CB\ by ultra-deep exposures with modern X-ray telescopes.
If the diffuse non-thermal X-ray emission is significantly below the upper limits derived in this paper, 
then the association of the $\gamma$-ray emitter with the radio lobes should be viewed as unlikely. 
Such a low level of the X-ray non-thermal halo would imply that the lobes are dominated by magnetic pressure, contrary to previous claims presented in the literature, or at least $\eta_{\rm eq} \simeq 1$.
This is because the X-ray--to--radio luminosity ratio is determined by the magnetic field energy density (or pressure) within the lobes, $L_{\rm X} / L_{\rm R} = U_{\rm cmb} / U_{\rm B} \propto B^{-2}$, where $U_{\rm cmb}$ is the energy density of the CMB photon field. 
The lower X-ray flux for a given radio flux implies a lower value of the $\eta_{\rm eq}$ parameter. 
In particular, noting that the inverse-Compton X-ray luminosity $L_{\rm X} \propto U_{\rm cmb} \times U_e \propto \eta_{\rm eq} \, U_{\rm B}$, one may find $\eta_{\rm eq} \propto L_{\rm X}^2$ for the fixed $L_{\rm R}$ and $U_{\rm cmb}$. 
In the modeling presented above, the expected X-ray luminosity of the lobes is only a factor of $\simeq 1.4$ below the provided \SZ\ upper limit, and the corresponding equipartition parameter is $\eta_{\rm eq} = 4$. 
On the other hand, if the X-ray flux is a factor of 3 below the current upper limit, the equipartition parameter is $\eta_{\rm eq} < 1$ and magnetic pressure dominates over the pressure of ultrarelativistic electrons within the lobes of \CB.

\section{Conclusions}

In this paper, we present a detailed analysis of about 43 months accumulation of \FL\ data and of newly acquired \SZ\ X-ray data for \CB. 
We detect the source in GeV photon energies at $4.5\,\sigma$ significance (${\rm TS}=29$).
The implied $\gamma$-ray ($0.2-200$\,GeV) luminosity of the system is $L_{\gamma} \simeq 6 \times 10^{42}$\,erg\,s$^{-1}$, and the observed continuum within the LAT range is best modeled by a power-law function with the photon index $\Gamma_{\gamma} = 2.6 \pm 0.2$. At X-ray frequencies we detect the core of \CB\ with \SZ\ at the flux level consistent with the previous {\MOD measurements with \emph{ASCA}, XMM-\emph{Newton} and \emph{Chandra}.} The unabsorbed $2-10$\,keV luminosity of the core is $L_{\rm X} \simeq 2 \times 10^{42}$\,erg\,s$^{-1}$, and the spectrum is best fitted with the power-law model with photon index $\Gamma_{\rm X} \simeq 1.62 \pm 0.05$. {\MOD We do not detect any significant diffuse X-ray emission of the lobes,} with the provided upper limit $F_{\rm 2-10\,keV} < 8.7 \times 10^{-13}$\,erg\,cm$^{-2}$\,s$^{-1}$, assuming the photon index $\Gamma_{\rm X}=1.7$. The upper limit is marginally consistent with the flux found by \citet{tashiro98} in their analysis of the \emph{ASCA} data, $F_{\rm 2-10\,keV} \simeq (9.8 \pm 3.8) \times 10^{-13}$\,erg\,cm$^{-2}$\,s$^{-1}$. 

By means of broad-band modeling we show that the observed $\gamma$-ray flux of the source may in principle be produced within the lobes via the inverse-Compton scattering of the CMB photon field by radio-emitting electrons, if the diffuse non-thermal X-ray emission component is not significantly below the derived \SZ\ upper limit.
This association would imply efficient \emph{in-situ} acceleration of ultrarelativistic particles within the lobes dominated by the particle pressure, in direct analogy to two other radio galaxies resolved by \FL, namely Centaurus\,A \citep{CenALobes} and NGC\,6251 \citep{takeuchi12}. 
However, if the diffuse X-ray emission in the studied object is much below the derived \SZ\ upper limits, then the observed $\gamma$-ray flux is not likely to be produced within the lobes, but instead within the unresolved core of the \CB\ radio galaxy. 
Assuming the one-zone synchrotron self-Compton model, we show that this possibility is also justified by the broad-band data collected for the unresolved core of Centaurus B, including the newly derived \SZ\ spectrum, and the expected parameters of the misaligned nuclear relativistic jet.
Interestingly, in such a case the extended lobes in the system could be dominated by the pressure of the magnetic field.

\begin{acknowledgements}
We thank M.~Ajello, S.~W.~Digel, J.~Finke, and Y.~Uchiyama for useful discussions.
We also appreciate the referee for his/her valuable comments, which improve our paper.
YT and JS acknowledge support from the Faculty of the European Space Astronomy Centre (ESAC).
\L .S. is grateful for the support from Polish MNiSW through the grant N-N203-380336.
Work by C.~C.~C.~at NRL is sponsored by NASA DPR S-15633-Y.

The \textit{Fermi} LAT Collaboration acknowledges generous ongoing support
from a number of agencies and institutes that have supported both the
development and the operation of the LAT as well as scientific data analysis.
These include the National Aeronautics and Space Administration and the
Department of Energy in the United States, the Commissariat \`a l'Energie Atomique
and the Centre National de la Recherche Scientifique / Institut National de Physique
Nucl\'eaire et de Physique des Particules in France, the Agenzia Spaziale Italiana
and the Istituto Nazionale di Fisica Nucleare in Italy, the Ministry of Education,
Culture, Sports, Science and Technology (MEXT), High Energy Accelerator Research
Organization (KEK) and Japan Aerospace Exploration Agency (JAXA) in Japan, and
the K.~A.~Wallenberg Foundation, the Swedish Research Council and the
Swedish National Space Board in Sweden.

Additional support for science analysis during the operations phase is gratefully
acknowledged from the Istituto Nazionale di Astrofisica in Italy and the Centre National d'\'Etudes Spatiales in France.
\end{acknowledgements}

\bibliographystyle{aa}
\bibliography{bibs_cenB}

\begin{thebibliography}{70}
\expandafter\ifx\csname natexlab\endcsname\relax\def\natexlab#1{#1}\fi

\bibitem[{{Abdo} {et~al.}(2009{\natexlab{a}}){Abdo}, {Ackermann}, {Ajello},
  {Asano}, {Baldini}, {Ballet}, {Barbiellini}, {Bastieri}, {Baughman},
  {Bechtol}, {Bellazzini}, {Blandford}, {Bloom}, {Bonamente}, {Borgland},
  {Bregeon}, {Brez}, {Brigida}, {Bruel}, {Burnett}, {Caliandro}, {Cameron},
  {Caraveo}, {Casandjian}, {Cavazzuti}, {Cecchi}, {Celotti}, {Chekhtman},
  {Cheung}, {Chiang}, {Ciprini}, {Claus}, {Cohen-Tanugi}, {Colafrancesco},
  {Cominsky}, {Conrad}, {Costamante}, {Dermer}, {de Angelis}, {de Palma},
  {Digel}, {Donato}, {do Couto e Silva}, {Drell}, {Dubois}, {Dumora},
  {Farnier}, {Favuzzi}, {Finke}, {Focke}, {Frailis}, {Fukazawa}, {Funk},
  {Fusco}, {Gargano}, {Georganopoulos}, {Germani}, {Giebels}, {Giglietto},
  {Giordano}, {Glanzman}, {Grenier}, {Grondin}, {Grove}, {Guillemot},
  {Guiriec}, {Hanabata}, {Harding}, {Hartman}, {Hayashida}, {Hays}, {Hughes},
  {J{\'o}hannesson}, {Johnson}, {Johnson}, {Johnson}, {Kadler}, {Kamae},
  {Kanai}, {Katagiri}, {Kataoka}, {Kawai}, {Kerr}, {Kn{\"o}dlseder}, {Kuehn},
  {Kuss}, {Latronico}, {Lemoine-Goumard}, {Longo}, {Loparco}, {Lott},
  {Lovellette}, {Lubrano}, {Madejski}, {Makeev}, {Mazziotta}, {McEnery},
  {Meurer}, {Michelson}, {Mitthumsiri}, {Mizuno}, {Moiseev}, {Monte},
  {Monzani}, {Morselli}, {Moskalenko}, {Murgia}, {Nakamori}, {Nolan}, {Norris},
  {Nuss}, {Ohsugi}, {Omodei}, {Orlando}, {Ormes}, {Paneque}, {Panetta},
  {Parent}, {Pepe}, {Pesce-Rollins}, {Piron}, {Porter}, {Rain{\`o}}, {Razzano},
  {Reimer}, {Reimer}, {Reposeur}, {Ritz}, {Rodriguez}, {Romani}, {Ryde},
  {Sadrozinski}, {Sambruna}, {Sanchez}, {Sander}, {Sato}, {Parkinson},
  {Sgr{\`o}}, {Smith}, {Smith}, {Spandre}, {Spinelli}, {Starck}, {Strickman},
  {Strong}, {Suson}, {Tajima}, {Takahashi}, {Takahashi}, {Tanaka}, {Taylor},
  {Thayer}, {Thompson}, {Torres}, {Tosti}, {Uchiyama}, {Usher}, {Vilchez},
  {Vitale}, {Waite}, {Wood}, {Ylinen}, {Ziegler}, {Aller}, {Aller},
  {Kellermann}, {Kovalev}, {Kovalev}, {Lister}, \& {Pushkarev}}]{PerA}
{Abdo}, A.~A., {Ackermann}, M., {Ajello}, M., {et~al.} 2009{\natexlab{a}},
  \apj, 699, 31

\bibitem[{{Abdo} {et~al.}(2009{\natexlab{b}}){Abdo}, {Ackermann}, {Ajello},
  {Atwood}, {Axelsson}, {Baldini}, {Ballet}, {Barbiellini}, {Bastieri},
  {Bechtol}, {Bellazzini}, {Berenji}, {Blandford}, {Bloom}, {Bonamente},
  {Borgland}, {Bregeon}, {Brez}, {Brigida}, {Bruel}, {Burnett}, {Caliandro},
  {Cameron}, {Cannon}, {Caraveo}, {Casandjian}, {Cavazzuti}, {Cecchi}, {{\c
  C}elik}, {Charles}, {Cheung}, {Chiang}, {Ciprini}, {Claus}, {Cohen-Tanugi},
  {Colafrancesco}, {Conrad}, {Costamante}, {Cutini}, {Davis}, {Dermer}, {de
  Angelis}, {de Palma}, {Digel}, {Donato}, {Silva}, {Drell}, {Dubois},
  {Dumora}, {Edmonds}, {Farnier}, {Favuzzi}, {Fegan}, {Finke}, {Focke},
  {Fortin}, {Frailis}, {Fukazawa}, {Funk}, {Fusco}, {Gargano}, {Gasparrini},
  {Gehrels}, {Georganopoulos}, {Germani}, {Giebels}, {Giglietto}, {Giommi},
  {Giordano}, {Giroletti}, {Glanzman}, {Godfrey}, {Grenier}, {Grondin},
  {Grove}, {Guillemot}, {Guiriec}, {Hanabata}, {Harding}, {Hayashida}, {Hays},
  {Horan}, {J{\'o}hannesson}, {Johnson}, {Johnson}, {Johnson}, {Johnson},
  {Kamae}, {Katagiri}, {Kataoka}, {Kawai}, {Kerr}, {Kn{\"o}dlseder}, {Kocian},
  {Kuss}, {Lande}, {Latronico}, {Lemoine-Goumard}, {Longo}, {Loparco}, {Lott},
  {Lovellette}, {Lubrano}, {Madejski}, {Makeev}, {Mazziotta}, {McConville},
  {McEnery}, {Meurer}, {Michelson}, {Mitthumsiri}, {Mizuno}, {Moiseev},
  {Monte}, {Monzani}, {Morselli}, {Moskalenko}, {Murgia}, {Nolan}, {Norris},
  {Nuss}, {Ohsugi}, {Omodei}, {Orlando}, {Ormes}, {Ozaki}, {Paneque},
  {Panetta}, {Parent}, {Pelassa}, {Pepe}, {Pesce-Rollins}, {Piron}, {Porter},
  {Rain{\`o}}, {Rando}, {Razzano}, {Reimer}, {Reimer}, {Reposeur}, {Ritz},
  {Rochester}, {Rodriguez}, {Romani}, {Roth}, {Ryde}, {Sadrozinski},
  {Sambruna}, {Sanchez}, {Sander}, {Saz Parkinson}, {Scargle}, {Sgr{\`o}},
  {Shaw}, {Smith}, {Smith}, {Spandre}, {Spinelli}, {Strickman}, {Suson},
  {Tajima}, {Takahashi}, {Tanaka}, {Taylor}, {Thayer}, {Thompson}, {Tibaldo},
  {Torres}, {Tosti}, {Tramacere}, {Uchiyama}, {Usher}, {Vasileiou}, {Vilchez},
  {Waite}, {Wang}, {Winer}, {Wood}, {Ylinen}, {Ziegler}, {Harris}, {Massaro},
  \& {Stawarz}}]{M87}
{Abdo}, A.~A., {Ackermann}, M., {Ajello}, M., {et~al.} 2009{\natexlab{b}},
  \apj, 707, 55

\bibitem[{{Abdo} {et~al.}(2010{\natexlab{a}}){Abdo}, {Ackermann}, {Ajello},
  {Atwood}, {Baldini}, {Ballet}, {Barbiellini}, {Bastieri}, {Baughman},
  {Bechtol}, {Bellazzini}, {Berenji}, {Blandford}, {Bloom}, {Bonamente},
  {Borgland}, {Bouvier}, {Brandt}, {Bregeon}, {Brez}, {Brigida}, {Bruel},
  {Buehler}, {Buson}, {Caliandro}, {Cameron}, {Cannon}, {Caraveo}, {Carrigan},
  {Casandjian}, {Cavazzuti}, {Cecchi}, {{\c C}elik}, {Charles}, {Chekhtman},
  {Cheung}, {Chiang}, {Ciprini}, {Claus}, {Cohen-Tanugi}, {Colafrancesco},
  {Cominsky}, {Conrad}, {Costamante}, {Davis}, {Dermer}, {de Angelis}, {de
  Palma}, {Silva}, {Drell}, {Dubois}, {Dumora}, {Falcone}, {Farnier},
  {Favuzzi}, {Fegan}, {Finke}, {Focke}, {Fortin}, {Frailis}, {Fukazawa},
  {Funk}, {Fusco}, {Gargano}, {Gasparrini}, {Gehrels}, {Georganopoulos},
  {Germani}, {Giebels}, {Giglietto}, {Giommi}, {Giordano}, {Giroletti},
  {Glanzman}, {Godfrey}, {Grandi}, {Grenier}, {Grondin}, {Grove}, {Guillemot},
  {Guiriec}, {Hadasch}, {Harding}, {Hase}, {Hayashida}, {Hays}, {Horan},
  {Hughes}, {Itoh}, {Jackson}, {J{\'o}hannesson}, {Johnson}, {Johnson},
  {Johnson}, {Kadler}, {Kamae}, {Katagiri}, {Kataoka}, {Kawai}, {Kishishita},
  {Kn{\"o}dlseder}, {Kuss}, {Lande}, {Latronico}, {Lee}, {Lemoine-Goumard},
  {Llena Garde}, {Longo}, {Loparco}, {Lott}, {Lovellette}, {Lubrano}, {Makeev},
  {Mazziotta}, {McConville}, {McEnery}, {Michelson}, {Mitthumsiri}, {Mizuno},
  {Moiseev}, {Monte}, {Monzani}, {Morselli}, {Moskalenko}, {Murgia},
  {M{\"u}ller}, {Nakamori}, {Naumann-Godo}, {Nolan}, {Norris}, {Nuss}, {Ohno},
  {Ohsugi}, {Ojha}, {Okumura}, {Omodei}, {Orlando}, {Ormes}, {Ozaki}, {Pagani},
  {Paneque}, {Panetta}, {Parent}, {Pelassa}, {Pepe}, {Pesce-Rollins}, {Piron},
  {Pl{\"o}tz}, {Porter}, {Rain{\`o}}, {Rando}, {Razzano}, {Razzaque}, {Reimer},
  {Reimer}, {Reposeur}, {Ripken}, {Ritz}, {Rodriguez}, {Roth}, {Ryde},
  {Sadrozinski}, {Sanchez}, {Sander}, {Scargle}, {Sgr{\`o}}, {Siskind},
  {Smith}, {Spandre}, {Spinelli}, {Starck}, {Stawarz}, {Strickman}, {Suson},
  {Tajima}, {Takahashi}, {Takahashi}, {Tanaka}, {Thayer}, {Thayer}, {Thompson},
  {Tibaldo}, {Torres}, {Tosti}, {Tramacere}, {Uchiyama}, {Usher},
  {Vandenbroucke}, {Vasileiou}, {Vilchez}, {Vitale}, {Waite}, {Wang}, {Winer},
  {Wood}, {Yang}, {Ylinen}, \& {Ziegler}}]{CenACore}
{Abdo}, A.~A., {Ackermann}, M., {Ajello}, M., {et~al.} 2010{\natexlab{a}},
  \apj, 719, 1433

\bibitem[{{Abdo} {et~al.}(2010{\natexlab{b}}){Abdo}, {Ackermann}, {Ajello},
  {Atwood}, {Baldini}, {Ballet}, {Barbiellini}, {Bastieri}, {Baughman},
  {Bechtol}, {Bellazzini}, {Berenji}, {Blandford}, {Bloom}, {Bonamente},
  {Borgland}, {Bregeon}, {Brez}, {Brigida}, {Bruel}, {Burnett}, {Buson},
  {Caliandro}, {Cameron}, {Caraveo}, {Casandjian}, {Cavazzuti}, {Cecchi}, {{\c
  C}elik}, {Chekhtman}, {Cheung}, {Chiang}, {Ciprini}, {Claus}, {Cohen-Tanugi},
  {Colafrancesco}, {Cominsky}, {Conrad}, {Costamante}, {Cutini}, {Davis},
  {Dermer}, {de Angelis}, {de Palma}, {Digel}, {do Couto e Silva}, {Drell},
  {Dubois}, {Dumora}, {Farnier}, {Favuzzi}, {Fegan}, {Finke}, {Focke},
  {Fortin}, {Fukazawa}, {Funk}, {Fusco}, {Gargano}, {Gasparrini}, {Gehrels},
  {Georganopoulos}, {Germani}, {Giebels}, {Giglietto}, {Giordano}, {Giroletti},
  {Glanzman}, {Godfrey}, {Grenier}, {Grove}, {Guillemot}, {Guiriec},
  {Hanabata}, {Harding}, {Hayashida}, {Hays}, {Hughes}, {Jackson},
  {J{\'o}hannesson G.}, {Johnson}, {Johnson}, {Johnson}, {Kamae}, {Katagiri},
  {Kataoka}, {Kawai}, {Kerr}, {Kn{\"o}dlseder}, {Kocian}, {Kuss}, {Lande},
  {Latronico}, {Lemoine-Goumard}, {Longo}, {Loparco}, {Lott}, {Lovellette},
  {Lubrano}, {Madejski}, {Makeev}, {Mazziotta}, {McConville}, {McEnery},
  {Meurer}, {Michelson}, {Mitthumsiri}, {Mizuno}, {Moiseev}, {Monte},
  {Monzani}, {Morselli}, {Moskalenko}, {Murgia}, {Nolan}, {Norris}, {Nuss},
  {Ohsugi}, {Omodei}, {Orlando}, {Ormes}, {Paneque}, {Parent}, {Pelassa},
  {Pepe}, {Pesce-Rollins}, {Piron}, {Porter}, {Rain{\`o}}, {Rando}, {Razzano},
  {Razzaque}, {Reimer}, {Reimer}, {Reposeur}, {Ritz}, {Rochester}, {Rodriguez},
  {Romani}, {Roth}, {Ryde}, {Sadrozinski}, {Sambruna}, {Sanchez}, {Sander},
  {Saz Parkinson}, {Scargle}, {Sgr{\`o}}, {Siskind}, {Smith}, {Smith},
  {Spandre}, {Spinelli}, {Starck}, {Stawarz}, {Strickman}, {Suson}, {Tajima},
  {Takahashi}, {Takahashi}, {Tanaka}, {Thayer}, {Thayer}, {Thompson},
  {Tibaldo}, {Torres}, {Tosti}, {Tramacere}, {Uchiyama}, {Vasileiou},
  {Vilchez}, {Vitale}, {Waite}, {Wallace}, {Wang}, {Winer}, {Wood}, {Ylinen},
  {Ziegler}, {Hardcastle}, {Kazanas}, \& {Fermi-LAT Collaboration}}]{CenALobes}
{Abdo}, A.~A., {Ackermann}, M., {Ajello}, M., {et~al.} 2010{\natexlab{b}},
  Science, 328, 725

\bibitem[{{Abdo} {et~al.}(2010{\natexlab{c}}){Abdo}, {Ackermann}, {Ajello},
  {Baldini}, {Ballet}, {Barbiellini}, {Bastieri}, {Bechtol}, {Bellazzini},
  {Berenji}, {Blandford}, {Bloom}, {Bonamente}, {Borgland}, {Bouvier},
  {Brandt}, {Bregeon}, {Brez}, {Brigida}, {Bruel}, {Buehler}, {Burnett},
  {Buson}, {Caliandro}, {Cameron}, {Cannon}, {Caraveo}, {Carrigan},
  {Casandjian}, {Cavazzuti}, {Cecchi}, {{\c C}elik}, {Celotti}, {Charles},
  {Chekhtman}, {Chen}, {Cheung}, {Chiang}, {Ciprini}, {Claus}, {Cohen-Tanugi},
  {Colafrancesco}, {Conrad}, {Davis}, {Dermer}, {de Angelis}, {de Palma},
  {Silva}, {Drell}, {Dubois}, {Favuzzi}, {Fegan}, {Ferrara}, {Fortin},
  {Frailis}, {Fukazawa}, {Fusco}, {Gargano}, {Gasparrini}, {Gehrels},
  {Germani}, {Giglietto}, {Giommi}, {Giordano}, {Giroletti}, {Glanzman},
  {Godfrey}, {Grandi}, {Grenier}, {Grove}, {Guillemot}, {Guiriec}, {Hadasch},
  {Hayashida}, {Hays}, {Horan}, {Hughes}, {Jackson}, {J{\'o}hannesson},
  {Johnson}, {Johnson}, {Kamae}, {Katagiri}, {Kataoka}, {Kn{\"o}dlseder},
  {Kuss}, {Lande}, {Latronico}, {Lee}, {Lemoine-Goumard}, {Llena Garde},
  {Longo}, {Loparco}, {Lott}, {Lovellette}, {Lubrano}, {Madejski}, {Makeev},
  {Malaguti}, {Mazziotta}, {McConville}, {McEnery}, {Michelson}, {Migliori},
  {Mitthumsiri}, {Mizuno}, {Monte}, {Monzani}, {Morselli}, {Moskalenko},
  {Murgia}, {Naumann-Godo}, {Nestoras}, {Nolan}, {Norris}, {Nuss}, {Ohsugi},
  {Okumura}, {Omodei}, {Orlando}, {Ormes}, {Paneque}, {Panetta}, {Parent},
  {Pelassa}, {Pepe}, {Persic}, {Pesce-Rollins}, {Piron}, {Porter}, {Rain{\`o}},
  {Rando}, {Razzano}, {Razzaque}, {Reimer}, {Reimer}, {Reyes}, {Roth},
  {Sadrozinski}, {Sanchez}, {Sander}, {Scargle}, {Sgr{\`o}}, {Siskind},
  {Smith}, {Spandre}, {Spinelli}, {Stawarz}, {Stecker}, {Strickman}, {Suson},
  {Takahashi}, {Tanaka}, {Thayer}, {Thayer}, {Thompson}, {Tibaldo}, {Torres},
  {Torresi}, {Tosti}, {Tramacere}, {Uchiyama}, {Usher}, {Vandenbroucke},
  {Vasileiou}, {Vilchez}, {Villata}, {Vitale}, {Waite}, {Wang}, {Winer},
  {Wood}, {Yang}, {Ylinen}, \& {Ziegler}}]{MAGN}
{Abdo}, A.~A., {Ackermann}, M., {Ajello}, M., {et~al.} 2010{\natexlab{c}},
  \apj, 720, 912

\bibitem[{{Abraham} {et~al.}(2007){Abraham}, {Abreu}, {Aglietta}, {Aguirre},
  {Allard}, {Allekotte}, {Allen}, {Allison}, {Alvarez}, \& et~al.}]{PAO}
{Abraham}, J., {Abreu}, P., {Aglietta}, M., {et~al.} 2007, Science, 318, 938

\bibitem[{{Abramowski} {et~al.}(2012){Abramowski}, {Acero}, {Aharonian},
  {Akhperjanian}, {Anton}, {Balzer}, {Barnacka}, {Barres de Almeida},
  {Becherini}, {Becker}, \& et~al.}]{M87MWL}
{Abramowski}, A., {Acero}, F., {Aharonian}, F., {et~al.} 2012, \apj, 746, 151

\bibitem[{{Ackermann} {et~al.}(2011){Ackermann}, {Ajello}, {Allafort},
  {Antolini}, {Atwood}, {Axelsson}, {Baldini}, {Ballet}, {Barbiellini},
  {Bastieri}, {Bechtol}, {Bellazzini}, {Berenji}, {Blandford}, {Bloom},
  {Bonamente}, {Borgland}, {Bottacini}, {Bouvier}, {Bregeon}, {Brigida},
  {Bruel}, {Buehler}, {Burnett}, {Buson}, {Caliandro}, {Cameron}, {Caraveo},
  {Casandjian}, {Cavazzuti}, {Cecchi}, {Charles}, {Cheung}, {Chiang},
  {Ciprini}, {Claus}, {Cohen-Tanugi}, {Conrad}, {Costamante}, {Cutini}, {de
  Angelis}, {de Palma}, {Dermer}, {Digel}, {Silva}, {Drell}, {Dubois},
  {Escande}, {Favuzzi}, {Fegan}, {Ferrara}, {Finke}, {Focke}, {Fortin},
  {Frailis}, {Fukazawa}, {Funk}, {Fusco}, {Gargano}, {Gasparrini}, {Gehrels},
  {Germani}, {Giebels}, {Giglietto}, {Giommi}, {Giordano}, {Giroletti},
  {Glanzman}, {Godfrey}, {Grenier}, {Grove}, {Guiriec}, {Gustafsson},
  {Hadasch}, {Hayashida}, {Hays}, {Healey}, {Horan}, {Hou}, {Hughes},
  {Iafrate}, {J{\'o}hannesson}, {Johnson}, {Johnson}, {Kamae}, {Katagiri},
  {Kataoka}, {Kn{\"o}dlseder}, {Kuss}, {Lande}, {Larsson}, {Latronico},
  {Longo}, {Loparco}, {Lott}, {Lovellette}, {Lubrano}, {Madejski}, {Mazziotta},
  {McConville}, {McEnery}, {Michelson}, {Mitthumsiri}, {Mizuno}, {Moiseev},
  {Monte}, {Monzani}, {Moretti}, {Morselli}, {Moskalenko}, {Murgia},
  {Nakamori}, {Naumann-Godo}, {Nolan}, {Norris}, {Nuss}, {Ohno}, {Ohsugi},
  {Okumura}, {Omodei}, {Orienti}, {Orlando}, {Ormes}, {Ozaki}, {Paneque},
  {Parent}, {Pesce-Rollins}, {Pierbattista}, {Piranomonte}, {Piron}, {Pivato},
  {Porter}, {Rain{\`o}}, {Rando}, {Razzano}, {Razzaque}, {Reimer}, {Reimer},
  {Ritz}, {Rochester}, {Romani}, {Roth}, {Sanchez}, {Sbarra}, {Scargle},
  {Schalk}, {Sgr{\`o}}, {Shaw}, {Siskind}, {Spandre}, {Spinelli}, {Strong},
  {Suson}, {Tajima}, {Takahashi}, {Takahashi}, {Tanaka}, {Thayer}, {Thayer},
  {Thompson}, {Tibaldo}, {Tinivella}, {Torres}, {Tosti}, {Troja}, {Uchiyama},
  {Vandenbroucke}, {Vasileiou}, {Vianello}, {Vitale}, {Waite}, {Wallace},
  {Wang}, {Winer}, {Wood}, {Wood}, \& {Zimmer}}]{2LAC}
{Ackermann}, M., {Ajello}, M., {Allafort}, A., {et~al.} 2011, \apj, 743, 171

\bibitem[{{Ackermann} {et~al.}(2012)}]{PASS7}
{Ackermann}, M. {et~al.} 2012, arXiv:1206.1896

\bibitem[{{Alexander}(2000)}]{alexander00}
{Alexander}, P. 2000, \mnras, 319, 8

\bibitem[{Atwood {et~al.}(2009)}]{LAT}
Atwood, W.~B. {et~al.} 2009, {\rm ApJ}, 697, 1071

\bibitem[{{B{\^i}rzan} {et~al.}(2008){B{\^i}rzan}, {McNamara}, {Nulsen},
  {Carilli}, \& {Wise}}]{Birzan08}
{B{\^i}rzan}, L., {McNamara}, B.~R., {Nulsen}, P.~E.~J., {Carilli}, C.~L., \&
  {Wise}, M.~W. 2008, \apj, 686, 859

\bibitem[{{Burke-Spolaor} {et~al.}(2009){Burke-Spolaor}, {Ekers}, {Massardi},
  {Murphy}, {Partridge}, {Ricci}, \& {Sadler}}]{burke2009}
{Burke-Spolaor}, S., {Ekers}, R.~D., {Massardi}, M., {et~al.} 2009, MNRAS, 395,
  504

\bibitem[{{Condon} {et~al.}(1993){Condon}, {Griffith}, \& {Wright}}]{pmn1993}
{Condon}, J.~J., {Griffith}, M.~R., \& {Wright}, A.~E. 1993, AJ, 106, 1095

\bibitem[{{Croston} {et~al.}(2005){Croston}, {Hardcastle}, {Harris}, {Belsole},
  {Birkinshaw}, \& {Worrall}}]{croston05}
{Croston}, J.~H., {Hardcastle}, M.~J., {Harris}, D.~E., {et~al.} 2005, \apj,
  626, 733

\bibitem[{{Croston} {et~al.}(2009){Croston}, {Kraft}, {Hardcastle},
  {Birkinshaw}, {Worrall}, {Nulsen}, {Penna}, {Sivakoff}, {Jord{\'a}n},
  {Brassington}, {Evans}, {Forman}, {Gilfanov}, {Goodger}, {Harris}, {Jones},
  {Juett}, {Murray}, {Raychaudhury}, {Sarazin}, {Voss}, \&
  {Woodley}}]{Croston09}
{Croston}, J.~H., {Kraft}, R.~P., {Hardcastle}, M.~J., {et~al.} 2009, \mnras,
  395, 1999

\bibitem[{{Donato} {et~al.}(2004){Donato}, {Sambruna}, \& {Gliozzi}}]{DD04}
{Donato}, D., {Sambruna}, R.~M., \& {Gliozzi}, M. 2004, \apj, 617, 915

\bibitem[{{Ebeling} {et~al.}(2002){Ebeling}, {Mullis}, \& {Tully}}]{ebeling02}
{Ebeling}, H., {Mullis}, C.~R., \& {Tully}, R.~B. 2002, \apj, 580, 774

\bibitem[{{Eracleous} {et~al.}(2000){Eracleous}, {Sambruna}, \&
  {Mushotzky}}]{eracleous00}
{Eracleous}, M., {Sambruna}, R., \& {Mushotzky}, R.~F. 2000, \apj, 537, 654

\bibitem[{{Evans} {et~al.}(2006){Evans}, {Worrall}, {Hardcastle}, {Kraft}, \&
  {Birkinshaw}}]{evans06}
{Evans}, D.~A., {Worrall}, D.~M., {Hardcastle}, M.~J., {Kraft}, R.~P., \&
  {Birkinshaw}, M. 2006, \apj, 642, 96

\bibitem[{{Feigelson} {et~al.}(1995){Feigelson}, {Laurent-Muehleisen},
  {Kollgaard}, \& {Fomalont}}]{feigelson95}
{Feigelson}, E.~D., {Laurent-Muehleisen}, S.~A., {Kollgaard}, R.~I., \&
  {Fomalont}, E.~B. 1995, \apjl, 449, L149

\bibitem[{{Fey} {et~al.}(2004){Fey}, {Ojha}, {Reynolds}, {Ellingsen},
  {McCulloch}, {Jauncey}, \& {Johnston}}]{Fey04}
{Fey}, A.~L., {Ojha}, R., {Reynolds}, J.~E., {et~al.} 2004, \aj, 128, 2593

\bibitem[{{Gopal-Krishna} \& {Wiita}(2000)}]{GK00}
{Gopal-Krishna} \& {Wiita}, P.~J. 2000, \aap, 363, 507

\bibitem[{{Hardcastle} {et~al.}(2009{\natexlab{a}}){Hardcastle}, {Cheung},
  {Feain}, \& {Stawarz}}]{hardcastle09}
{Hardcastle}, M.~J., {Cheung}, C.~C., {Feain}, I.~J., \& {Stawarz}, {\L}.
  2009{\natexlab{a}}, \mnras, 393, 1041

\bibitem[{{Hardcastle} {et~al.}(2009{\natexlab{b}}){Hardcastle}, {Evans}, \&
  {Croston}}]{HEC09}
{Hardcastle}, M.~J., {Evans}, D.~A., \& {Croston}, J.~H. 2009{\natexlab{b}},
  \mnras, 396, 1929

\bibitem[{{Ishisaki} {et~al.}(2007){Ishisaki}, {Maeda}, {Fujimoto}, {Ozaki},
  {Ebisawa}, {Takahashi}, {Ueda}, {Ogasaka}, {Ptak}, {Mukai}, {Hamaguchi},
  {Hirayama}, {Kotani}, {Kubo}, {Shibata}, {Ebara}, {Furuzawa}, {Iizuka},
  {Inoue}, {Mori}, {Okada}, {Yokoyama}, {Matsumoto}, {Nakajima}, {Yamaguchi},
  {Anabuki}, {Tawa}, {Nagai}, {Katsuda}, {Hayashida}, {Bamba}, {Miller},
  {Sato}, \& {Yamasaki}}]{XISSIM}
{Ishisaki}, Y., {Maeda}, Y., {Fujimoto}, R., {et~al.} 2007, \pasj, 59, 113

\bibitem[{{Isobe} {et~al.}(2011){Isobe}, {Seta}, \& {Tashiro}}]{isobe11}
{Isobe}, N., {Seta}, H., \& {Tashiro}, M.~S. 2011, \pasj, 63, 947

\bibitem[{{Isobe} {et~al.}(2009){Isobe}, {Tashiro}, {Gandhi}, {Hayato},
  {Nagai}, {Hada}, {Seta}, \& {Matsuta}}]{isobe09}
{Isobe}, N., {Tashiro}, M.~S., {Gandhi}, P., {et~al.} 2009, \apj, 706, 454

\bibitem[{{Jones} {et~al.}(2001){Jones}, {Lloyd}, \& {McAdam}}]{JLM01}
{Jones}, P.~A., {Lloyd}, B.~D., \& {McAdam}, W.~B. 2001, {\rm MNRAS}, 325, 817

\bibitem[{{Kaneda} {et~al.}(1995){Kaneda}, {Tashiro}, {Ikebe}, {Ishisaki},
  {Kubo}, {Makshima}, {Ohashi}, {Saito}, {Tabara}, \& {Takahashi}}]{kaneda95}
{Kaneda}, H., {Tashiro}, M., {Ikebe}, Y., {et~al.} 1995, \apjl, 453, L13

\bibitem[{{Kataoka} \& {Stawarz}(2005)}]{kataoka05}
{Kataoka}, J. \& {Stawarz}, {\L}. 2005, \apj, 622, 797

\bibitem[{{Kataoka} {et~al.}(2010){Kataoka}, {Stawarz}, {Cheung}, {Tosti},
  {Cavazzuti}, {Celotti}, {Nishino}, {Fukazawa}, {Thompson}, \&
  {McConville}}]{kataoka10}
{Kataoka}, J., {Stawarz}, {\L}., {Cheung}, C.~C., {et~al.} 2010, \apj, 715, 554

\bibitem[{{Kataoka} {et~al.}(2011){Kataoka}, {Stawarz}, {Takahashi}, {Cheung},
  {Hayashida}, {Grandi}, {Burnett}, {Celotti}, {Fegan}, {Fortin}, {Maeda},
  {Nakamori}, {Taylor}, {Tosti}, {Digel}, {McConville}, {Finke}, \&
  {D'Ammando}}]{BLRG}
{Kataoka}, J., {Stawarz}, {\L}., {Takahashi}, Y., {et~al.} 2011, \apj, 740, 29

\bibitem[{{Kawakatu} {et~al.}(2008){Kawakatu}, {Nagai}, \& {Kino}}]{kawakatu08}
{Kawakatu}, N., {Nagai}, H., \& {Kino}, M. 2008, \apj, 687, 141

\bibitem[{{Konar} {et~al.}(2006){Konar}, {Saikia}, {Jamrozy}, \&
  {Machalski}}]{konar06}
{Konar}, C., {Saikia}, D.~J., {Jamrozy}, M., \& {Machalski}, J. 2006, \mnras,
  372, 693

\bibitem[{{Koyama} {et~al.}(2007){Koyama}, {Tsunemi}, {Dotani}, {Bautz},
  {Hayashida}, {Tsuru}, {Matsumoto}, {Ogawara}, {Ricker}, {Doty}, {Kissel},
  {Foster}, {Nakajima}, {Yamaguchi}, {Mori}, {Sakano}, {Hamaguchi},
  {Nishiuchi}, {Miyata}, {Torii}, {Namiki}, {Katsuda}, {Matsuura}, {Miyauchi},
  {Anabuki}, {Tawa}, {Ozaki}, {Murakami}, {Maeda}, {Ichikawa}, {Prigozhin},
  {Boughan}, {Lamarr}, {Miller}, {Burke}, {Gregory}, {Pillsbury}, {Bamba},
  {Hiraga}, {Senda}, {Katayama}, {Kitamoto}, {Tsujimoto}, {Kohmura}, {Tsuboi},
  \& {Awaki}}]{Koyama2007}
{Koyama}, K., {Tsunemi}, H., {Dotani}, T., {et~al.} 2007, \pasj, 59, 23

\bibitem[{{Kraan-Korteweg} \& {Lahav}(2000)}]{kraan00}
{Kraan-Korteweg}, R.~C. \& {Lahav}, O. 2000, \aapr, 10, 211

\bibitem[{{Kraan-Korteweg} {et~al.}(2005){Kraan-Korteweg}, {Ochoa}, {Woudt}, \&
  {Andernach}}]{kraan05}
{Kraan-Korteweg}, R.~C., {Ochoa}, M., {Woudt}, P.~A., \& {Andernach}, H. 2005,
  in Astronomical Society of the Pacific Conference Series, Vol. 329, Nearby
  Large-Scale Structures and the Zone of Avoidance, ed. {A.~P.~Fairall \&
  P.~A.~Woudt}, 159--165

\bibitem[{{Kraan-Korteweg} \& {Woudt}(1999)}]{kraan99}
{Kraan-Korteweg}, R.~C. \& {Woudt}, P.~A. 1999, PASA, 16, 53

\bibitem[{{Laing} \& {Bridle}(2002)}]{laing02}
{Laing}, R.~A. \& {Bridle}, A.~H. 2002, \mnras, 336, 1161

\bibitem[{{Laing} {et~al.}(2011){Laing}, {Guidetti}, {Bridle}, {Parma}, \&
  {Bondi}}]{Laing11}
{Laing}, R.~A., {Guidetti}, D., {Bridle}, A.~H., {Parma}, P., \& {Bondi}, M.
  2011, \mnras, 417, 2789

\bibitem[{{Laing} {et~al.}(1999){Laing}, {Parma}, {de Ruiter}, \&
  {Fanti}}]{laing99}
{Laing}, R.~A., {Parma}, P., {de Ruiter}, H.~R., \& {Fanti}, R. 1999, \mnras,
  306, 513

\bibitem[{{Laustsen} {et~al.}(1977){Laustsen}, {Schuster}, \&
  {West}}]{laustsen77}
{Laustsen}, S., {Schuster}, H.-E., \& {West}, R.~M. 1977, \aap, 59, L3

\bibitem[{{Lynden-Bell} {et~al.}(1988){Lynden-Bell}, {Faber}, {Burstein},
  {Davies}, {Dressler}, {Terlevich}, \& {Wegner}}]{GA}
{Lynden-Bell}, D., {Faber}, S.~M., {Burstein}, D., {et~al.} 1988, \apj, 326, 19

\bibitem[{{Machalski} {et~al.}(2007){Machalski}, {Chy{\.z}y}, {Stawarz}, \&
  {Kozie{\l}}}]{machalski07}
{Machalski}, J., {Chy{\.z}y}, K.~T., {Stawarz}, {\L}., \& {Kozie{\l}}, D. 2007,
  \aap, 462, 43

\bibitem[{{Marshall} {et~al.}(2005){Marshall}, {Schwartz}, {Lovell}, {Murphy},
  {Worrall}, {Birkinshaw}, {Gelbord}, {Perlman}, \& {Jauncey}}]{marshall05}
{Marshall}, H.~L., {Schwartz}, D.~A., {Lovell}, J.~E.~J., {et~al.} 2005, \apjs,
  156, 13

\bibitem[{Mattox {et~al.}(1996)}]{Mattox96}
Mattox, J.~R. {et~al.} 1996, {\rm ApJ}, 461, 396

\bibitem[{{McAdam}(1991)}]{mcadam1991}
{McAdam}, W.~B. 1991, Proceedings of the Astronomical Society of Australia, 9,
  255

\bibitem[{{Mitsuda} {et~al.}(2007){Mitsuda}, {Bautz}, {Inoue}, {Kelley},
  {Koyama}, {Kunieda}, {Makishima}, {Ogawara}, {Petre}, {Takahashi}, {Tsunemi},
  {White}, {Anabuki}, {Angelini}, {Arnaud}, {Awaki}, {Bamba}, {Boyce}, {Brown},
  {Chan}, {Cottam}, {Dotani}, {Doty}, {Ebisawa}, {Ezoe}, {Fabian}, {Figueroa},
  {Fujimoto}, {Fukazawa}, {Furusho}, {Furuzawa}, {Gendreau}, {Griffiths},
  {Haba}, {Hamaguchi}, {Harrus}, {Hasinger}, {Hatsukade}, {Hayashida}, {Henry},
  {Hiraga}, {Holt}, {Hornschemeier}, {Hughes}, {Hwang}, {Ishida}, {Ishisaki},
  {Isobe}, {Itoh}, {Iyomoto}, {Kahn}, {Kamae}, {Katagiri}, {Kataoka},
  {Katayama}, {Kawai}, {Kilbourne}, {Kinugasa}, {Kissel}, {Kitamoto}, {Kohama},
  {Kohmura}, {Kokubun}, {Kotani}, {Kotoku}, {Kubota}, {Madejski}, {Maeda},
  {Makino}, {Markowitz}, {Matsumoto}, {Matsumoto}, {Matsuoka}, {Matsushita},
  {McCammon}, {Mihara}, {Misaki}, {Miyata}, {Mizuno}, {Mori}, {Mori}, {Morii},
  {Moseley}, {Mukai}, {Murakami}, {Murakami}, {Mushotzky}, {Nagase}, {Namiki},
  {Negoro}, {Nakazawa}, {Nousek}, {Okajima}, {Ogasaka}, {Ohashi}, {Oshima},
  {Ota}, {Ozaki}, {Ozawa}, {Parmar}, {Pence}, {Porter}, {Reeves}, {Ricker},
  {Sakurai}, {Sanders}, {Senda}, {Serlemitsos}, {Shibata}, {Soong}, {Smith},
  {Suzuki}, {Szymkowiak}, {Takahashi}, {Tamagawa}, {Tamura}, {Tamura},
  {Tanaka}, {Tashiro}, {Tawara}, {Terada}, {Terashima}, {Tomida}, {Torii},
  {Tsuboi}, {Tsujimoto}, {Tsuru}, {Turner}, {Ueda}, {Ueno}, {Ueno}, {Uno},
  {Urata}, {Watanabe}, {Yamamoto}, {Yamaoka}, {Yamasaki}, {Yamashita},
  {Yamauchi}, {Yamauchi}, {Yaqoob}, {Yonetoku}, \& {Yoshida}}]{Mitsuda2007}
{Mitsuda}, K., {Bautz}, M., {Inoue}, H., {et~al.} 2007, \pasj, 59, 1

\bibitem[{{Moskalenko} {et~al.}(2009){Moskalenko}, {Stawarz}, {Porter}, \&
  {Cheung}}]{moskalenko09}
{Moskalenko}, I.~V., {Stawarz}, {\L}., {Porter}, T.~A., \& {Cheung}, C.~C.
  2009, \apj, 693, 1261

\bibitem[{{Nagar} \& {Matulich}(2008)}]{nagar08}
{Nagar}, N.~M. \& {Matulich}, J. 2008, \aap, 488, 879

\bibitem[{{Nagayama} {et~al.}(2004){Nagayama}, {Woudt}, {Nagashima},
  {Nakajima}, {Kato}, {Kurita}, {Nagata}, {Nakaya}, {Tamura}, {Sugitani},
  {Wakamatsu}, \& {Sato}}]{nagayama04}
{Nagayama}, T., {Woudt}, P.~A., {Nagashima}, C., {et~al.} 2004, \mnras, 354,
  980

\bibitem[{{Nolan} {et~al.}(2012){Nolan}, {Abdo}, {Ackermann}, {Ajello},
  {Allafort}, {Antolini}, {Atwood}, {Axelsson}, {Baldini}, {Ballet}, \&
  et~al.}]{2FGL}
{Nolan}, P.~L., {Abdo}, A.~A., {Ackermann}, M., {et~al.} 2012, \apjs, 199, 31

\bibitem[{{O'Sullivan} {et~al.}(2009){O'Sullivan}, {Reville}, \&
  {Taylor}}]{osullivan09}
{O'Sullivan}, S., {Reville}, B., \& {Taylor}, A.~M. 2009, \mnras, 400, 248

\bibitem[{{Pe'er} \& {Loeb}(2012)}]{peer12}
{Pe'er}, A. \& {Loeb}, A. 2012, JCAP, 3, 7

\bibitem[{{Planck Collaboration} {et~al.}(2011{\natexlab{a}}){Planck
  Collaboration}, {Ade}, {Aghanim}, {Arnaud}, {Ashdown}, {Aumont},
  {Baccigalupi}, {Baker}, {Balbi}, {Banday}, \& et~al.}]{Planck}
{Planck Collaboration}, {Ade}, P.~A.~R., {Aghanim}, N., {et~al.}
  2011{\natexlab{a}}, \aap, 536, A1

\bibitem[{{Planck Collaboration} {et~al.}(2011{\natexlab{b}}){Planck
  Collaboration}, {Ade}, {Aghanim}, {Arnaud}, {Ashdown}, {Aumont},
  {Baccigalupi}, {Balbi}, {Banday}, {Barreiro}, \& et~al.}]{PlanckERCSC}
{Planck Collaboration}, {Ade}, P.~A.~R., {Aghanim}, N., {et~al.}
  2011{\natexlab{b}}, \aap, 536, A7

\bibitem[{{Raue} \& {Mazin}(2008)}]{RM08}
{Raue}, M. \& {Mazin}, D. 2008, International Journal of Modern Physics D, 17,
  1515

\bibitem[{{Revnivtsev} {et~al.}(2006){Revnivtsev}, {Sazonov}, {Gilfanov},
  {Churazov}, \& {Sunyaev}}]{Revnivtsev2006}
{Revnivtsev}, M., {Sazonov}, S., {Gilfanov}, M., {Churazov}, E., \& {Sunyaev},
  R. 2006, \aap, 452, 169

\bibitem[{{Scheuer}(1995)}]{scheuer95}
{Scheuer}, P.~A.~G. 1995, \mnras, 277, 331

\bibitem[{{Schr{\"o}der} {et~al.}(2007){Schr{\"o}der}, {Mamon},
  {Kraan-Korteweg}, \& {Woudt}}]{schroeder07}
{Schr{\"o}der}, A.~C., {Mamon}, G.~A., {Kraan-Korteweg}, R.~C., \& {Woudt},
  P.~A. 2007, \aap, 466, 481

\bibitem[{{Serlemitsos} {et~al.}(2007){Serlemitsos}, {Soong}, {Chan},
  {Okajima}, {Lehan}, {Maeda}, {Itoh}, {Mori}, {Iizuka}, {Itoh}, {Inoue},
  {Okada}, {Yokoyama}, {Itoh}, {Ebara}, {Nakamura}, {Suzuki}, {Ishida},
  {Hayakawa}, {Inoue}, {Okuma}, {Kubota}, {Suzuki}, {Osawa}, {Yamashita},
  {Kunieda}, {Tawara}, {Ogasaka}, {Furuzawa}, {Tamura}, {Shibata}, {Haba},
  {Naitou}, \& {Misaki}}]{Serlemitsos2007}
{Serlemitsos}, P.~J., {Soong}, Y., {Chan}, K.-W., {et~al.} 2007, \pasj, 59, 9

\bibitem[{{Siemiginowska} {et~al.}(2012){Siemiginowska}, {Stawarz}, {Cheung},
  {Aldcroft}, {Bechtold}, {Burke}, {Evans}, {Holt}, {Jamrozy}, \&
  {Migliori}}]{Siemiginowska12}
{Siemiginowska}, A., {Stawarz}, {\L}., {Cheung}, C.~C., {et~al.} 2012, \apj,
  750, 124

\bibitem[{{Svoboda} {et~al.}(2012){Svoboda}, {Bianchi}, {Guainazzi}, {Matt},
  {Piconcelli}, {Karas}, \& {Dov{\v c}iak}}]{4U1344}
{Svoboda}, J., {Bianchi}, S., {Guainazzi}, M., {et~al.} 2012, \aap, 545, A148

\bibitem[{{Takeuchi} {et~al.}(2012){Takeuchi}, {Kataoka}, {Stawarz},
  {Takahashi}, {Maeda}, {Nakamori}, {Cheung}, {Celotti}, {Tanaka}, \&
  {Takahashi}}]{takeuchi12}
{Takeuchi}, Y., {Kataoka}, J., {Stawarz}, L., {et~al.} 2012, ArXiv e-prints

\bibitem[{{Tashiro} {et~al.}(2005){Tashiro}, {Isobe}, {Suzuki}, {Ito}, {Abe},
  \& {Makishima}}]{tashiro05}
{Tashiro}, M., {Isobe}, N., {Suzuki}, M., {et~al.} 2005, in X-Ray and Radio
  Connections, ed. {L.~O.~Sjouwerman \& K.~K.~Dyer}

\bibitem[{{Tashiro} {et~al.}(1998){Tashiro}, {Kaneda}, {Makishima}, {Iyomoto},
  {Idesawa}, {Ishisaki}, {Kotani}, {Takahashi}, \& {Yamashita}}]{tashiro98}
{Tashiro}, M., {Kaneda}, H., {Makishima}, K., {et~al.} 1998, \apj, 499, 713

\bibitem[{{Tawa} {et~al.}(2008){Tawa}, {Hayashida}, {Nagai}, {Nakamoto},
  {Tsunemi}, {Yamaguchi}, {Ishisaki}, {Miller}, {Mizuno}, {Dotani}, {Ozaki}, \&
  {Katayama}}]{Tawa08}
{Tawa}, N., {Hayashida}, K., {Nagai}, M., {et~al.} 2008, \pasj, 60, 11

\bibitem[{{West} \& {Tarenghi}(1989)}]{west89}
{West}, R.~M. \& {Tarenghi}, M. 1989, \aap, 223, 61

\bibitem[{{Yuasa} {et~al.}(2009){Yuasa}, {Nakazawa}, \& {Makishima}}]{Yuasa09}
{Yuasa}, T., {Nakazawa}, K., \& {Makishima}, K. 2009, \pasj, 61, 1107

\end{thebibliography}

\end{document}